\documentclass[preprint2]{aastex6}
\usepackage{color}
\usepackage[titletoc]{appendix}
\usepackage[fleqn]{amsmath}
\usepackage{mathtools}
\usepackage{float}
\usepackage{comment}
\usepackage{enumitem}
\usepackage{natbib}
\usepackage{bm}
\usepackage{totcount}

\newtotcounter{citnum} %From the package documentation
\def\oldbibitem{} \let\oldbibitem=\bibitem
\def\bibitem{\stepcounter{citnum}\oldbibitem}

\shortauthors{Millholland}
\shorttitle{Obliquity Tides in Sub-Neptunes}

\begin{document} 

%This document contains \total{citnum}\ references.

\title{Tidally-Induced Radius Inflation of Sub-Neptunes} 
\author{Sarah Millholland}
\affil{Department of Astronomy, Yale University, New Haven, CT 06511, USA}
\altaffiliation{NSF Graduate Research Fellow}
\email{sarah.millholland@yale.edu}

\begin{abstract}
Recent work suggests that many short-period super-Earth and sub-Neptune planets may have significant spin axis tilts (``obliquities''). When planets are locked in high-obliquity states, the tidal dissipation rate may increase by several orders of magnitude. This intensified heat deposition within the planets' interiors should generate significant structural consequences, including atmospheric inflation leading to larger transit radii. Using up-to-date radius estimates from \textit{Gaia} Data Release 2, we show evidence for $\sim50\%$ larger average radii of planets wide of first-order mean-motion resonances, a population of planets with a theorized frequent occurrence of high obliquities. We investigate whether this radius trend could be a signature of obliquity tides. Using an adaptation of the Modules for Experiments in Stellar Astrophysics (MESA) stellar evolution toolkit, we model the atmospheric evolution of sub-Neptune-mass planets in response to additional internal heat from obliquity tides. The degree of radius inflation predicted by the models is $\sim10\%-100\%$ for tidal luminosities $\gtrsim 10^{-5}$ of the incident stellar power; this degree of inflation is broadly consistent with the observations and can approximately be described by power law relationships. We present a few case studies of very low density ``super-puff'' planets -- Kepler-79 d, Kepler-31 c, and Kepler-27 b -- and show that they are strong candidates for potentially having undergone tidally-induced radius inflation. We also discuss how the discrepancy between the two populations of planets with masses derived from radial velocities and transit timing variations is connected to the radius distribution features we have identified. Altogether, the calculations in this work confirm that tidal dissipation has non-negligible consequences for the structural properties of short-period sub-Neptunes.
%While it is still uncertain whether obliquity tides are in fact the cause of these features, this work confirms the non-negligible consequences of tidal dissipation on the structural properties of short-period sub-Neptunes. 
\end{abstract}

\section{Introduction}

NASA's \textit{Kepler} mission fundamentally altered the paradigm and future outlook of exoplanet science when it showed that sub-Neptune-sized planets on short-period orbits ($P\lesssim 100 \ \mathrm{days}$) -- now often simply termed ``Kepler planets'' -- are ubiquitous \citep[e.g.][]{2011ApJ...736...19B, 2013ApJS..204...24B}. Roughly $\sim 30\%-50\%$ of Sun-like stars contain such planets \citep{2012ApJS..201...15H, 2013ApJ...766...81F, 2013PNAS..11019273P, 2015ARA&A..53..409W, 2018ApJ...860..101Z}. 

Although the \textit{Kepler} prime mission ended in 2013 and the spacecraft has been retired for nearly a year, our understanding of these pervasive yet unfamiliar worlds continues to advance. The recent spectroscopic California-\textit{Kepler} Survey \citep[CKS,][]{2017AJ....154..107P, 2017AJ....154..108J}
%which obtained spectroscopy of planet-hosting stars in the Kepler prime field
and the \textit{Gaia} mission all-sky astrometric survey \citep{2018A&A...616A...1G} have significantly improved the radius measurement precision of \textit{Kepler} host stars and therefore also the transiting planets that orbit them. Meanwhile, the usage of transit timing variations \citep[TTVs,][]{2005Sci...307.1288H, 2005MNRAS.359..567A} and high-precision radial velocity (RV) measurements \citep[e.g.][]{2014ApJ...783L...6W} continue to provide quality mass estimates.  

Simultaneous with the growth and improvement of the observations, there is an expanding literature of theory aimed at characterizing the interior structures, compositions, and atmospheric evolution of these planets. Inherent degeneracies prohibit full specification of sub-Neptune compositions from their masses and radii alone; the typical densities, which range from $\rho \sim 0.1-10 \ \mathrm{g \ cm^{-3}}$, are frequently consistent with the compositions of both ``gas dwarfs'' -- rocky cores surrounded by envelopes dominated by hydrogen and helium -- and ``water worlds'' -- planets dominated by H\textsubscript{2}O ices/fluids \citep[e.g.][]{2008ApJ...673.1160A, 2010ApJ...712..974R}. 

The favored interpretation is that these sub-Neptunes are gas dwarfs with rocky cores and H/He-dominated envelopes that are typically $\sim0.05\%-5\%$ (but sometimes $\gtrsim 10\%$) of their total mass \citep{2013ApJ...775...10V, 2013ApJ...775..105O, 2014ApJ...792....1L, 2015ApJ...801...41R, 2015ApJ...806..183W, 2016ApJ...831..180C, 2017ApJ...847...29O, 2019ApJ...874...91W}. A wealth of modeling efforts have improved our understanding of these planets' initial formation and accretion \citep[e.g.][]{2015MNRAS.448.1751I, 2015ApJ...811...41L, 2016ApJ...817...90L,  2016ApJ...825...29G}, mass loss from irradiation-driven photoevaporation \citep{2010A&A...516A..20V, 2012ApJ...761...59L, 2013ApJ...775..105O, 2013ApJ...776....2L, 2014ApJ...795...65J, 2015ApJ...808..150H} and from the planets' own cooling \citep{2018MNRAS.476..759G, 2019MNRAS.487...24G}, and overall evolution to their present-day structures \citep{2014ApJ...792....1L, 2016ApJ...831..180C, 2018ApJ...869..163V}.

% Vazan: https://iopscience.iop.org/article/10.3847/1538-4357/aaef33/pdf

% Benneke https://arxiv.org/pdf/1907.00449.pdf

%Paper: Adams et al. (2008): https://iopscience.iop.org/article/10.1086/524925/pdf

% Ginzburg, Schlichting, & Sari https://iopscience.iop.org/article/10.3847/0004-637X/825/1/29/pdf

In addition to the planet's mass, composition, and stellar irradiation, another influence on the physical structure of a close-in sub-Neptunes is its spin state. It is often assumed that planets with $P\lesssim 100 \ \mathrm{days}$ have zero axial tilt (``obliquity'') and spins that are synchronized due to tides raised on the planets from their host stars. The timescale for this to occur is $\lesssim 10^7-10^8$ yr for planets with $P\lesssim 100 \ \mathrm{days}$. Recent work suggests, however, that some (perhaps many) planets in short-period, compact, and nearly-coplanar systems may have significant obliquities due to their intrinsic proximity and resultant capture in secular spin-orbit resonances \citep{2019NatAs...3..424M}. These resonances can excite and maintain planetary obliquities at large values even in the presence of tidal dissipation, which is intensified by several orders of magnitude at high obliquity.
 
% https://arxiv.org/pdf/1903.01386.pdf
\cite{2019NatAs...3..424M} proposed that the enhanced tides at high obliquity (``obliquity tides'') could explain the statistical overabundance of systems with pairs of planets just wide of the 3:2 and 2:1 orbital period ratios, the first-order mean-motion resonances (MMRs). Previous theories had shown that this could be dissipation-related, likely tidal \citep{2012ApJ...756L..11L, 2013AJ....145....1B}, and \cite{2014A&A...570L...7D} found an orbital period dependence in the offsets from MMRs that provided stronger evidence for a tidal origin. However, eccentricity tides alone were found to be insufficiently strong to produce the observed effect \citep{2013ApJ...774...52L, 2015MNRAS.453.4089S}; the extra dissipation from obliquity tides could be the solution.

The obliquity tides theory of the wide-of-MMR pile-up results in two corollaries pertaining to planet structure \citep{2019NatAs...3..424M}. First, it leads to an estimation of the typical tidal quality factor of sub-Neptunes, $Q \sim 10^4$, similar to the dissipation efficiencies of Uranus and Neptune \citep{1990Icar...85..394T, 2008Icar..193..267Z}. Second, since the wide-of-MMR planets are thought to have preferentially high obliquities, it suggests that obliquity tides may generate observable radius inflation for planets in these states \citep[e.g.][]{2008ApJ...681.1631J}.

In this paper, we explore the latter idea. We show observational evidence that planets in pairs wide of first-order MMRs have larger average radii. We then employ a thermal evolution model to investigate whether this could be a sign of tidally-induced inflation. Although our specific application is the obliquity tides theory of the near-MMR pile-up, there is also a broader motivation to study the structural impacts of tidal heating, as this extends to any short-period sub-Neptune that maintains a significant eccentricity or obliquity. 

This paper is organized as follows: We start by examining the planet radius distribution in Section \ref{section 2} and showing evidence for an enhancement of radii of planets in pairs wide of first-order MMRs. In Section \ref{section 3}, we describe our thermal evolution model, which builds off an existing sub-Neptune model \citep{2016ApJ...831..180C} to include tidal heating. We apply the model in Section \ref{section 4} to characterize when, where, and how much tides can affect planet structures. In Section \ref{section 5}, we identify and investigate case studies that are strong candidates for having undergone tidal radius inflation. Section \ref{section 6} is a discussion of further implications of this work, most notably a new appraisal of the discrepancy between the populations of TTV and RV planets. Finally, we summarize and conclude in Section \ref{section 7}.

\section{Planets in Pairs Wide of First-order MMR have Larger Average Radii}
\label{section 2}

An important prediction results from the obliquity tides theory of the near-MMR pile-up \citep{2019NatAs...3..424M}: Planets in pairs with period ratios wide of MMR should have larger average radii compared to those just inside MMR. This effect was indeed found to be present and statistically significant in the planet population  \citep{2019NatAs...3..424M}. However, this earlier analysis of the radius distribution utilized data from the NASA Exoplanet Archive\footnote{\href{https://exoplanetarchive.ipac.caltech.edu}{https://exoplanetarchive.ipac.caltech.edu}}, and updated radius estimates are available.
%from both \textit{Gaia} DR2 and the California-\textit{Kepler} Survey (CKS). 
The significance of the effect should increase when using a dataset with updated precision. In this section, we show that this true, and we study the effect in greater detail. 

%We began by compiling all confirmed and candidate planets (4723 in total) from the \textit{Kepler} DR 25 KOI Catalog \citep{2018ApJS..235...38T} via the NASA Exoplanet Archive\footnote{\href{https://exoplanetarchive.ipac.caltech.edu}{https://exoplanetarchive.ipac.caltech.edu}}. Where available, we updated the radii of these planets using a catalog of revised parameters derived by \citep{2018ApJ...866...99B}, who combined \textit{Kepler} DR 25 with parallaxes from \textit{Gaia} DR2. For planets not in the \cite{2018ApJ...866...99B} catalog, we checked the CKS catalog. If neither \textit{Gaia} or CKS updates were available, we used the DR 25 parameters. Finally, we removed 94 planet candidates that didn't have radius constraints. 

We begin with a catalog of 4045 \textit{Kepler} confirmed and candidate planets from \cite{2018ApJ...866...99B}, who leveraged parallaxes from \textit{Gaia} DR2 to derive updated radii for these planets. We cross-match this catalog with the \textit{Kepler} DR 25 KOI Catalog \citep{2018ApJS..235...38T} obtained via the NASA Exoplanet Archive\footnotemark[1] and remove 120 objects that have false positive dispositions, resulting in 3925 confirmed and candidate planets with \textit{Gaia}-precision radii. We extract all multiple planet systems, which are then used to construct the distribution of period ratios of pairs of planets within these systems. 

Our goal is to compare the average radii of planets in pairs on either side of the first-order MMRs. However, because there is a slight positive trend between the size of a planet and its period ratio with an adjacent planet \citep{2018AJ....155...48W}, we must compare the radius ratios across the first-order MMRs to typical radius ratios throughout the whole period ratio distribution. To that end, let us calculate the mean planet radii within small bins in the period ratio distribution and take ratios of these mean radii across adjacent bins. We denote this 
\begin{equation}
\label{average radius ratio}
r(x_i) = \frac{\mathrm{mean}({R_p | x_i < x < x_{i+1})}}{\mathrm{mean}({R_p | x_{i-1} < x < x_i})}
\end{equation}
where $x$ is the period ratio of a planet pair and $x_i$ are the bin edges. We take 0.05 width bins and $x_i \in \{1, 1.05, 1.1, ..., 3.9, 3.95, 4\}$, such that bins across the exact 3:2 and 2:1 ratios are included. In addition, we replace the 1.35 bin edge by 1.33, such that two bins (albeit now of slightly unequal widths) bridge the exact 4:3 ratio. We obtain errors on these mean radius ratios using a bootstrapping calculation with 1000 samples of the distribution with replacement. We perform three different versions of this calculation: a version in which the radii of both the inner and outer planets in the pair are involved in the averaging (equation \ref{average radius ratio}), and versions using only the inner or outer planets at a time.

\begin{figure}
\epsscale{1.2}
\plotone{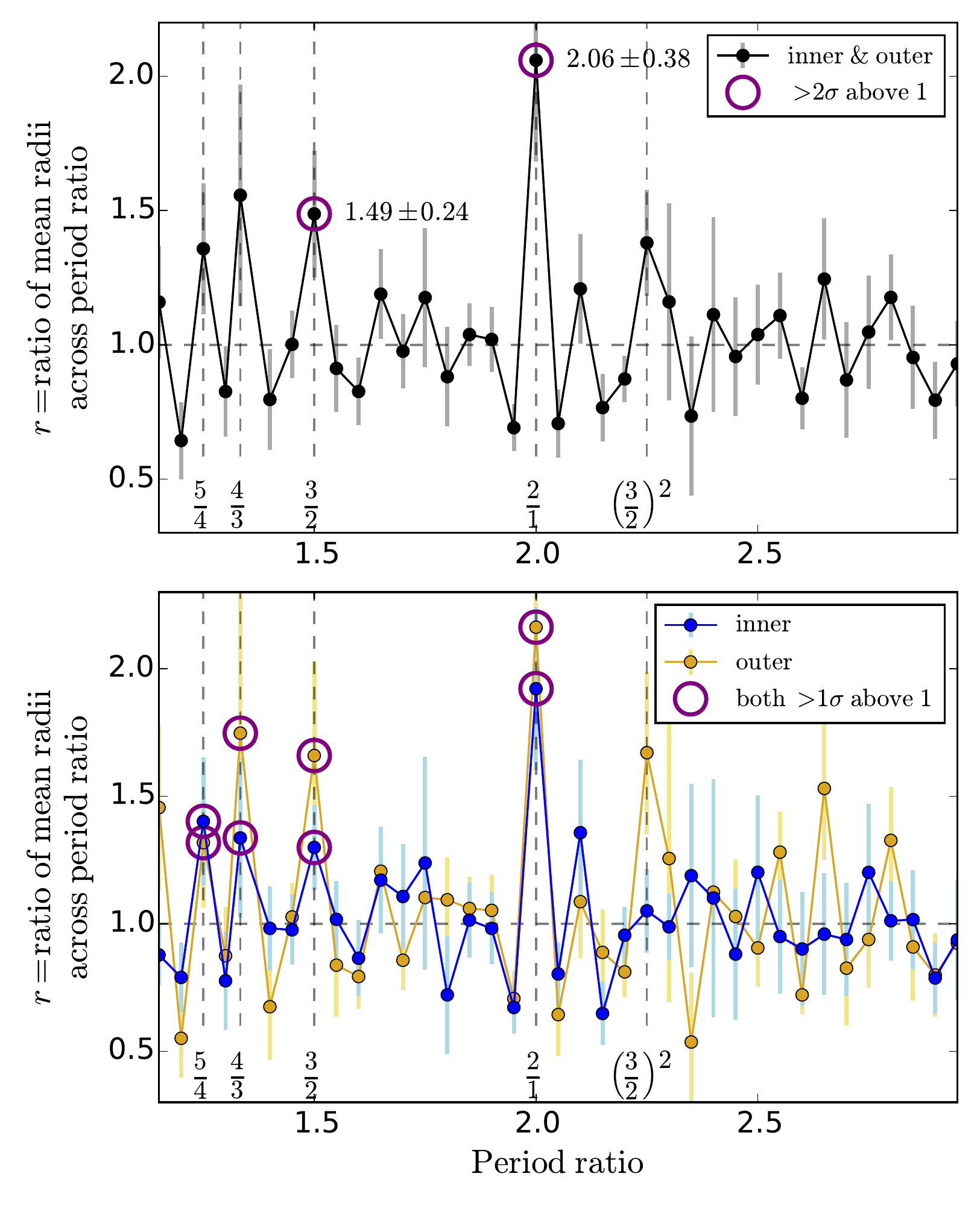}
\caption{The ratio, $r$, of mean radii across adjacent period ratio bins of width 0.05 (equation \ref{ratio_of_average_radii}). \textit{Top panel:} The mean is performed including both the inner and outer planets in a given period ratio pair, and the purple circles indicate values that are $> 2\sigma$ above unity. The values at $2/1$ and $3/2$ are labeled, though note that enhancements are seen at all first-order MMRs. \textit{Bottom panel:} The blue/yellow curves consider only inner/outer planets in the pairs, respectively, and the purple circles indicate values where both are $> 1\sigma$ above unity.}
\label{ratio_of_average_radii}
\end{figure}

The results of these calculations are shown in Figure \ref{ratio_of_average_radii}. It is clear to see that the ratios of mean radii across adjacent bins on either side of the first-order MMRs are enhanced; that is, $r(2/1)$ and $r(3/2)$, as well as $r(4/3)$ and $r(5/4)$, are all greater than unity. Table \ref{table 1} shows the values for $r(2/1)$ and $r(3/2)$. The planet sample involved in these near-MMR estimations is reasonably-sized; there are 62 and 31 pairs of planets in the bins just outside and inside the 3:2 MMR, respectively, and 37 and 10 pairs just outside and inside the 2:1 MMR. Using Mood's median test, the difference between radii in the bins of width 0.05 on either side of these MMRs is significant with a combined p-value of $p\approx10^{-4}$. The mean radius ratio is 1.70 when considering both the 3:2 and 2:1 MMRs simultaneously. 

Not only are the mean radius ratios across MMRs greater than 1; they are also enhanced by a greater degree than at any other period ratio. For the calculations using both the inner and outer planets (top panel of Figure \ref{ratio_of_average_radii} and first row of Table \ref{table 1}), $r(2/1)$ is $2.81\sigma$ above unity; the respective value for $r(3/2)$ is $2.06\sigma$. There are no other period ratios with a mean radius ratio $>2.0\sigma$ above unity; there is one that is close ($1.93\sigma$), and interestingly, that is $(3/2)^2 = 2.25$. This period ratio is close to $2.2$, a value at which \cite{2015MNRAS.448.1956S} noted a significant excess in the period ratio distribution. The peak in the period ratio distribution at 2.2, however, is probably not associated with the radius feature at 2.25; the latter may be related to resonant chains of 3:2 MMRs wherein the middle planet is non-transiting.

Examining the bottom panel of Figure \ref{ratio_of_average_radii}, which splits the calculation into only inner planets in the pairs and only outer planets in the pairs, we observe that the radius ratios for both inner and outer planets are simultaneously enhanced at all first-order MMRs. This simultaneous enhancement is not seen at any other period ratio. In addition, we observe that it is the outer planets that have an enhanced radius ratio at $(3/2)^2 = 2.25$.

In order to probe the strength of these observations and the extent to which they correspond to the inflation of gaseous envelopes, it is worthwhile to examine how the radius trends change when we limit the sizes of the planets in the sample. We repeat the earlier calculations, but this time, when calculating the mean radius ratio, we only consider planets with $R_p > 1.8 \ R_{\oplus}$. This is the approximate lower bound defining sub-Neptune planets \citep[e.g.][]{2018MNRAS.479.4786V}. The results for $r(2/1)$ and $r(3/2)$ are shown in Table \ref{table 1}. The mean radius ratios across these MMRs (as well as 5:4 and 4:3) are still enhanced, but they are less enhanced than the previous calculation. This is, in part, an expected effect of the smaller sample size and dynamic range that results when restricting the calculations to planets with $R_p > 1.8 \ R_{\oplus}$. Nevertheless, we suggest that this calculation is the more appropriate one to use if, as we hypothesize in this paper, the radius enhancement is due to tidal heating, since the calculation involves comparisons within populations of planets that are known to have gaseous components.

We summarize these observations as follows: (1) Planets in pairs wide of first-order MMRs have larger average radii compared to those just inside MMRs. (2) The observed radius difference is more than the typical degree of difference observed across other period ratios. (3) The effect is weaker but still apparent when restricting the sample to planets with $R_p > 1.8 \ R_{\oplus}$.

\begin{table}[h]
\centering
\caption{\textbf{Mean radius ratios.} The ratio, $r$, of mean radii across adjacent period ratio bins of width 0.05 on either side of the 2:1 and 3:2 MMRs (equation \ref{average radius ratio}). As in Figure \ref{ratio_of_average_radii}, the ratios are shown for three separate cases: inner and outer planet pairs together, only inner planets in the pair, and only outer planets in the pair. The middle column uses all planet radii; the final column uses $R_p > 1.8 \ R_{\oplus}$.}
\begin{tabular}{c | c | c}
 & All $R_p$ & $R_p > 1.8 \ R_{\oplus}$ \\
\hline
$r(2/1)$ inner \& outer & $2.06\pm0.38$ & $1.64\pm0.28$    \\
$r(2/1)$ inner only & $1.92\pm0.32$  & $1.47\pm0.15$  \\
$r(2/1)$ outer only & $2.16\pm0.53$ & $1.78\pm0.40$   \\
\hline
$r(3/2)$ inner \& outer & $1.49\pm0.24$  & $1.28\pm0.23$  \\
$r(3/2)$ inner only & $1.3\pm0.17$ & $0.98\pm0.16$   \\
$r(3/2)$ outer only & $1.66\pm0.37$ & $1.52\pm0.38$   \\
\end{tabular}
\label{table 1}
\end{table}

\subsection{Physical interpretations}
\label{section 2.1}
The hypothesis in this paper is that these radius distribution features are produced by tidal heating. However, this is not the only plausible explanation; there are additional factors to consider, and in reality, the observed radius enhancement may be a product of both tidal heating and formation and evolution conditions. 

First, planets with larger masses are more likely to be found in MMRs than those with smaller masses. This is because the probability of long-term capture into first-order resonance increases with the total planet to star mass ratio \citep{2014AJ....147...32G, 2015MNRAS.451.2589B, 2015ApJ...810..119D, 2019MNRAS.482..530T}. It is therefore expected that planets in or just wide of MMRs should have larger average \textit{masses} than those just inside. While this could be related to the radius trends, the mass-radius relation in the super-Earth/sub-Neptune regime is replete with intrinsic scatter due to compositional diversity \citep[e.g.][]{2016ApJ...825...19W}, so it is unclear that an enhancement in masses wide of MMRs would necessarily translate to the strong trends in radius. 

%Pan \& Schlichting: https://arxiv.org/pdf/1704.07836.pdf

In addition, \cite{2016ApJ...817...90L} postulated that ultra-low-density ``super-puff'' planets should preferentially be found in or near MMR chains after forming in the outer disk and migrating inwards. This effect may indeed contribute to the radius distribution signatures we have found. Similarly, the features could also be related to systematic differences in planetary core compositions. Though the formation of resonant chains may involve long-range migration, it likely does not require it \citep{2018AJ....156..228M}, so it is unclear how much of a role this is playing. 
% MacDonald: https://ui.adsabs.harvard.edu/abs/2018AJ....156..228M/abstract
%Lee & Chiang stuff about Super-puffs: Basically they argue that super-puffs form in the outer disk and migrate inwards, so they should preferentially be found in MMR chains.
%https://iopscience.iop.org/article/10.3847/0004-637X/817/2/90/meta
%The conundrum of super-Earths is that many of them have too little gas for their large core masses. The flip side of this puzzle is presented by "super-puffs": a rare class of Kepler planets characterized by orbital periods lesssim 50 days, radii ~4–10 R⊕, masses ~2–6 M⊕, and inferred GCRs gsim 20percent (e.g., Lopez & Fortney 2014; Masuda 2014). Super-puffs seem to have too much gas for their small core masses. To attain their large GCRs, super-puff cores must have enjoyed an environment that enabled rapid atmospheric cooling. In Section 4, we show how such an environment obtains at large stellocentric distances, outside ~1 AU—thus implicating orbital migration for the origin of super-puffs.

Regardless of the origin of this trend that wide-of-MMR planets have larger average radii, it is nonetheless useful to examine tides as a potential solution and, more generally, understand the structural implications of strong tidal dissipation in a sub-Neptune. It is expected that the planet's radius would be affected \citep[e.g.][]{2008ApJ...681.1631J}, but the degree of inflation is not immediately obvious. \cite{2017MNRAS.464.3937G} studied tidal heating in super-Earths/sub-Neptunes during the formation epoch and showed that tides could inhibit the rate of accretion of gas from the nebula. Our study is related but distinct in that it pertains to radius evolution on Gyr timescales rather than during the first few Myr. 

% Bodenheimer et al. (2001): https://ui.adsabs.harvard.edu/abs/2001ApJ...548..466B/abstract

Radius inflation of exoplanets has been studied almost entirely in the context of hot Jupiters, short-period ($P \lesssim 10$ days) giant planets whose sizes are infamously at odds with thermal evolution models \citep[e.g.][]{2002A&A...385..156G, 2003ApJ...592..555B, 2010RPPh...73a6901B, 2011ApJ...729L...7L}. These planets require the presence of some anomalous heating to account for their distended sizes \citep[e.g.][]{2018AJ....155..214T, 2018A&A...616A..76S, 2018ARA&A..56..175D}. Tidal dissipation has been one of many proposed origins of this extra heat \citep{2001ApJ...548..466B, 2004ApJ...608.1076G, 2004ApJ...610..477O, 2005ApJ...628L.159W, 2007A&A...462L...5L, 2007ApJ...665..754F, 2008ApJ...681.1631J, 2009ApJ...700.1921I, 2009ApJ...702.1413M, 2010ApJ...713..751I, 2010A&A...516A..64L}, but it is not the favored solution to explain the anomalous radii of all hot Jupiters. These studies have collectively characterized the inflationary behavior of hot Jupiters as they respond to extra heat with variable strength and deposition depth \citep[e.g.][]{2007ApJ...661..502B, 2015ApJ...803..111G, 2017ApJ...844...94K}. While the results for hot Jupiters do not directly translate to sub-Neptunes on account of their different structures, these studies can still aid in the interpretation of new modeling efforts.  

% Ibgui \& Burrows (2009): We make the default assumption of all planet radius evolutionary modelers to date that it is deposited in the core, but the reader is encouraged to keep an open mind... In our calculations, we assume that tidal heating occurs entirely in the convective interior of the planet and that the evolutionary process starts a few Myr after the star’s formation
% Cite Jackson papers and Gu et al. (2003)
% Cite Rasio et al. (1996)? \citep{1996ApJ...470.1187R}

% Sestovic et al. (2018): https://arxiv.org/pdf/1804.03075.pdf

% Thorngren et al. (2018): https://arxiv.org/pdf/1709.04539.pdf

% Jackson et al. (2008): ``Tidal Heating of Extrasolar Planets'' This one is a more general look at tidal heating on all exoplanets, not just hot Jupiters.

% Ibgui & Burrows (2009) https://ui.adsabs.harvard.edu/abs/2009ApJ...700.1921I/abstract
% Miller et al. (2009): https://ui.adsabs.harvard.edu/abs/2009ApJ...702.1413M/abstract
% Ibgui et al. (2010): https://ui.adsabs.harvard.edu/abs/2010ApJ...713..751I/abstract
% Leconte et al. (2010): https://ui.adsabs.harvard.edu/abs/2010A%26A...516A..64L/abstract
% Ogilvie & Lin: https://iopscience.iop.org/article/10.1086/421454/pdf

\newpage
\section{Thermal Evolution Model}
\label{section 3}

With the observation of larger average radii for planets in pairs wide of first-order MMRs as our primary motivation, we turn to examine the question of whether the obliquity tide hypothesis offers a consistent explanation. More generally, we wish to understand the impacts of tidal heating on a short-period planet's interior structure and atmosphere. To address this, we build onto the sub-Neptune evolutionary model developed by \cite{2016ApJ...831..180C}, which is robust, thoroughly-tested and benchmarked, and publicly available.

The \cite{2016ApJ...831..180C} model is an adaptation of the stellar evolution toolkit, Modules for Experiments in Stellar Astrophysics \citep[MESA;][]{2011ApJS..192....3P, 2013ApJS..208....4P, 2015ApJS..220...15P, 2018ApJS..234...34P}. In our implementation, we use MESA version 11554 and MESA SDK version 20190315. The model consists of a spherically symmetric planet with a heavy element core and an envelope dominated by hydrogen and helium. The H/He envelope is evolved using the one-dimensional stellar evolution \texttt{MESAstar} module using several MESA defaults. These include the H/He equation of state from \cite{1995ApJS...99..713S}, solar values of the metallicity (Z=0.03) and helium fraction (Y=0.25), and low-temperature Rosseland opacities from \cite{2008ApJS..174..504F, 2014ApJS..214...25F}. The planet is subject to irradiation flux from its host star, whose evolution is not modeled explicitly. (One can, however, model both the star and planet simultaneously using the \texttt{MESAbinary} module, as has been done by \citealt{2015ApJ...813..101V} for hot Jupiters.)

Prior to the work by \cite{2016ApJ...831..180C}, MESA already included functionality (described in \citealt{2013ApJS..208....4P}) designed to model planets down to masses as small as $\sim0.1M_{\mathrm{Jup}}$ \citep[e.g.][]{2013ApJ...763...13W, 2013ApJ...769L...9B, 2013ApJ...775..105O, 2014ApJ...793L...3V, 2015ApJ...813..101V, 2016CeMDA.126..227J}. In order to evolve highly-irradiated sub-Neptunes with masses down to $\sim 1 M_{\oplus}$, the model developed by \cite{2016ApJ...831..180C} employed several modifications to MESA. They included a new atmospheric boundary condition using the $T(\tau)$ relation of \cite{2010A&A...520A..27G},  hydrodynamic evaporative mass loss  driven by EUV and X-ray radiation from the stellar host using prescriptions from \cite{2009ApJ...693...23M}, and a physical model \citep{2011ApJ...738...59R} to determine the density and luminosity of the heavy-element core. In our implementation, we mostly use a ``rocky'' composition ($70\%$ silicates and $30\%$ Fe), but we consider other compositions in Section \ref{section 4.3}.

\subsection{Tidal heating}
\label{section 3.1}

We build upon the \cite{2016ApJ...831..180C} model by including extra dissipation in the planet's core. This dissipation is presumed to arise from tides raised on the planet from its host star; we do not model the gravitational tidal interactions or planetary deformation explicitly. Tides are fundamentally complex, and the details of where and how the energy is deposited are uncertain and non-trivial. For fully-formed sub-Neptunes, however, dissipation in the core should dominate over dissipation in the envelope. The ratio of the tidal power deposited in the envelope to that in the core is \citep{2017MNRAS.464.3937G}
\begin{equation}
\label{tidal power ratio}
\frac{P_{\mathrm{env}}}{P_{\mathrm{core}}} \approx \frac{M_{\mathrm{env}}}{M_{\mathrm{core}}}\frac{Q_{\mathrm{core}}}{Q_{\mathrm{env}}}\left(\frac{R_p}{R_{\mathrm{core}}}\right)^5.
\end{equation}
For $M_{\mathrm{env}}/M_{\mathrm{core}} = 0.01$, $Q_{\mathrm{core}} = 10^2$, $Q_{\mathrm{env}} = 10^5$, and $R_p/R_{\mathrm{core}} = 2$, this power ratio is $P_{\mathrm{env}}/P_{\mathrm{core}} \sim 10^{-4}$, indicating that dissipation in the core is significantly more dominant. Nonetheless, it is still important to understand how sensitively our results depend on our choice of deposition depth, so in Section \ref{section 4.4} we explore alternative heat deposition profiles.   

% Cite limitations, drawbacks, other models (e.g. Quillen) that would be more detailed. Though mroe detailed models might be useful, we know so little about tides so its unclear how'd they be useful. We just want an order-of-magnitude plausibility argument anyway!

The thermal evolution of a sub-Neptune is dominated by the luminosity of its core, since it makes up $\sim 90\%-99\%$ of the total mass. Within existing astrophysical models in the literature \citep[e.g.][]{2011ApJ...733....2N, 2012ApJ...761...59L, 2015ApJ...808..150H, 2016ApJ...831..180C, 2018A&A...610L...1V, 2018ApJ...869..163V}, 
% the thermal evolution of the core is not followed explicitly; instead, the core cooling rate, dTcore/dt is simply taken to be identical to the cooling at the base of the envelope
there are multiple core energy sources that are typically accounted for. %One source is the core's initial energy from formation and differentiation, which is accounted for as thermal inertia. A second source is heating from radioactive decay. Vazan et al. (also others?) have also considered the release of latent heat from solidification. 
The contribution from tidal dissipation has not yet been considered among these, but if the planet maintains an eccentric orbit or non-zero obliquity, this energy will be significant and sustained throughout the planet's lifetime, in contrast with some other core energy sources that only matter early on.
% Given the potentially significant and sustained contribution to the core energy, it is therefore worthwhile to explore the effect of tides on a short-period sub-Neptune's structure and thermal evolution.

Several sources of core luminosity, $L_{\mathrm{core}}$, are present in the \cite{2016ApJ...831..180C} model. We add a term for the tidal luminosity, such that $L_{\mathrm{core}}$ becomes  
\begin{equation}
L_{\mathrm{core}} = -c_v M_{\mathrm{core}}\frac{dT_{\mathrm{core}}}{dt} + L_{\mathrm{radio}} + L_{\mathrm{tide}}.
\end{equation}
The first term accounts for core cooling following initial formation and differentiation, where $c_v$ is the effective heat capacity at constant volume. The second term represents heating from decay of radioactive nuclides. Finally, the last term is the tidal luminosity, which we specify below. 

\begin{figure}
\epsscale{1.2}
\plotone{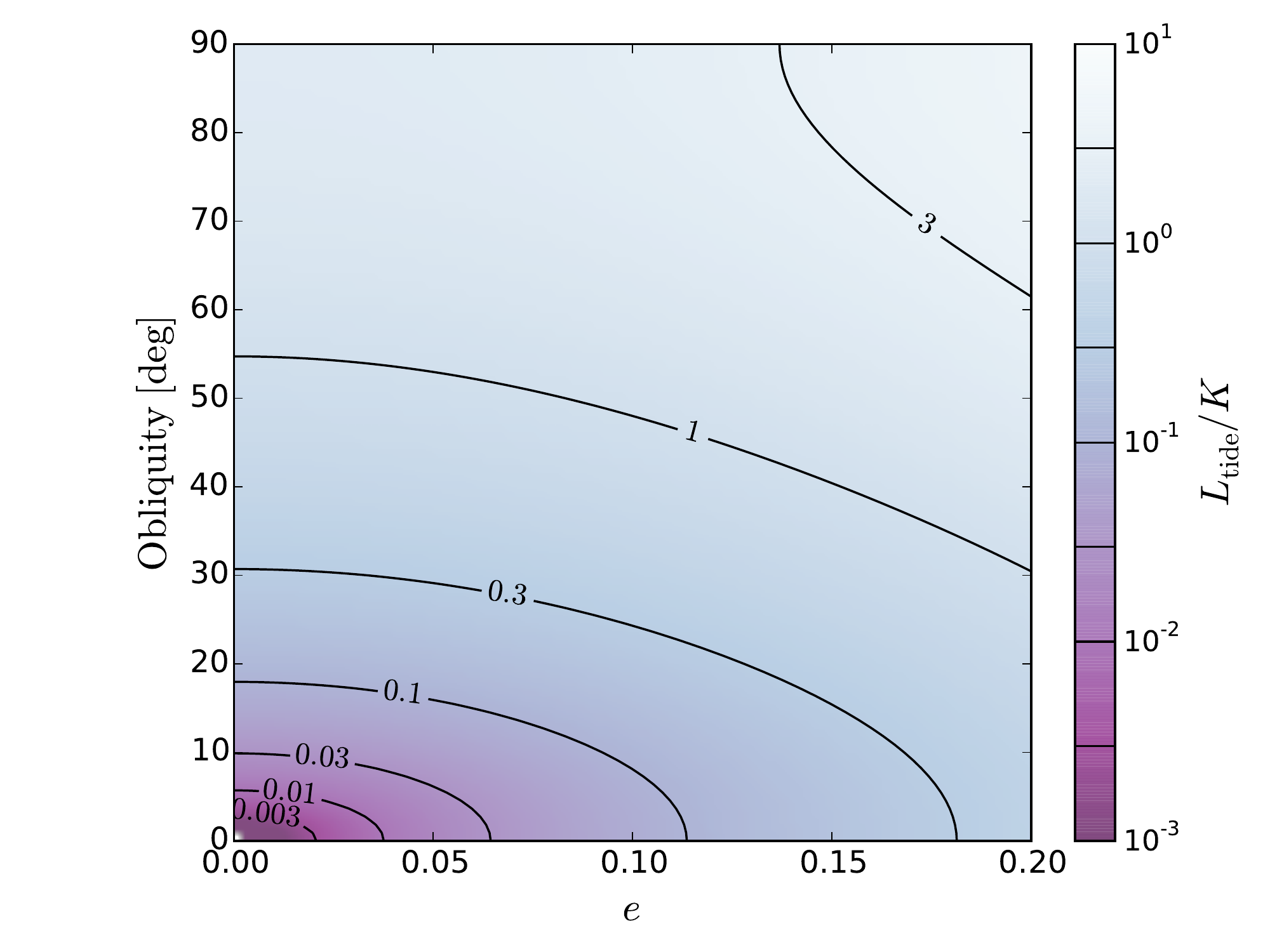}
\caption{The normalized tidal dissipation rate, $L_{\mathrm{tide}}/K$ (equation \ref{full dissipation rate}), as a function of eccentricity, $e$, and obliquity, $\epsilon$. At fixed eccentricity, the tidal dissipation rate increases by several orders of magnitude as $\epsilon$ increases from zero.} 
\label{Edot_over_K heatmap}
\end{figure}

We parameterize the tidal dissipation using the viscous approach to traditional equilibrium tide theory \citep{1880RSPT..171..713D, 1966Icar....5..375G, 1979M&P....20..301M, 1981A&A....99..126H, 2008Icar..193..637W, 2008CeMDA.101..171F, 2009CeMDA.104..257E, 2010A&A...516A..64L}. In this framework, the planet's tidal response to gravitational stresses from the star is an equilibrium deformation (tidal bulge). The bulge does not track the star's position exactly but rather lags with a time offset that, in one approach to the theory, may be approximated as constant. The equilibrium tide model effectively condenses the physics of the tidal distortion into a single parameter, $Q'$, the ``reduced tidal quality factor'', which will be further defined below. It is worth noting that there are many additional tidal models with more complex relations between the phase lag angle and the tidal forcing frequency \citep[e.g.][]{2009CeMDA.104..257E, 2012CeMDA.112..283E, 2013CeMDA.116..109F,  2014MNRAS.438.1526S, 2014A&A...571A..50C, 2016CeMDA.126...31B}, but given that the specific rheologies of sub-Neptunes are uncertain, we believe that the mathematically and physically simple approach of the equilibrium tide model is appropriate for our goal of deducing the first-order physical response.

Let us assume that the planet's rotation rate, $\omega$, has reached equilibrium (at which $d\omega/d t = 0$). The equilibrium rate is given by \citep{2007A&A...462L...5L}
\begin{equation}
\frac{\omega_{\mathrm{eq}}}{n} = \frac{N(e)}{\Omega(e)}\frac{2\cos\epsilon}{1+\cos^2\epsilon}
\label{omega eq}
\end{equation}
where $n=2\pi/P$ is the mean-motion, $e$ is the orbital eccentricity, $\epsilon$ is the planetary obliquity, and $N(e)$ and $\Omega(e)$ are functions of $e$ defined by 
\begin{align}
N(e) &= \frac{1 + \frac{15}{2}e^2 + \frac{45}{8}e^4 + \frac{5}{16}e^6}{(1-e^2)^6} \\
\Omega(e) &= \frac{1 + 3e^2 + \frac{3}{8}e^4}{(1-e^2)^{\frac{9}{2}}}.
\end{align}
The assumption that $\omega = \omega_{\mathrm{eq}}$ is appropriate for planets in the regime of our interest, since the timescale to reach this is $\omega/\dot{\omega} \lesssim 10^7 \ \mathrm{yr}$ for $a\lesssim0.3 \ \mathrm{AU}$ \citep[e.g.][]{2007A&A...462L...5L, 2017CeMDA.129..509B}.
% https://arxiv.org/pdf/1708.02981.pdf

With equilibrium rotation, the rate at which tidal dissipation converts orbital energy into heat energy is given by \citep{2007A&A...462L...5L} 
\begin{equation}
L_{\mathrm{tide}}(e,\epsilon) = 2 K\left[N_a(e) - \frac{N^2(e)}{\Omega(e)}\frac{2\cos^2\epsilon}{1+\cos^2\epsilon}\right] 
\label{full dissipation rate}
\end{equation}
with $N(e)$ and $\Omega(e)$ as defined above and
\begin{equation}
N_a(e) = \frac{1 + \frac{31}{2}e^2 + \frac{255}{8}e^4 + \frac{185}{16}e^6 + \frac{25}{64}e^8}{(1-e^2)^{\frac{15}{2}}}. \\
\end{equation}
To second order in eccentricity, $L_{\mathrm{tide}}$ is approximately 
\begin{equation}
L_{\mathrm{tide}}(e,\epsilon) = \frac{2K}{1+\cos^2\epsilon}[\sin^2\epsilon + e^2(7+16\sin^2\epsilon)].
\label{dissipation rate}
\end{equation}
The quantity $K$ in equations \ref{full dissipation rate} and \ref{dissipation rate} is
\begin{equation}
K = \frac{3n}{2}\frac{k_2}{Q}\left(\frac{G {M_{\star}}^2}{R_p}\right)\left(\frac{R_p}{a}\right)^6, 
\label{tidalK}
\end{equation}
where $M_{\star}$ is the stellar mass, $R_p$ the planet radius, and $a$ the semi-major axis. The two quantities $k_2$ and $Q$ are connected to the planet's composition and interior structure. The dimensionless Love number, $k_2$, is related to the central concentration of the planet's density profile and its deformation response to tidal disturbance. $Q = (n\Delta{t})^{-1}$ is the annual tidal quality factor (where $\Delta t$ is the constant tidal time lag), and it is related to the efficiency of tidal damping. It is custom to combine $Q$ and $k_2$ into the so-called ``reduced tidal quality factor'', $Q' = 3Q/2k_2$.  

\begin{figure}
\epsscale{1.2}
\plotone{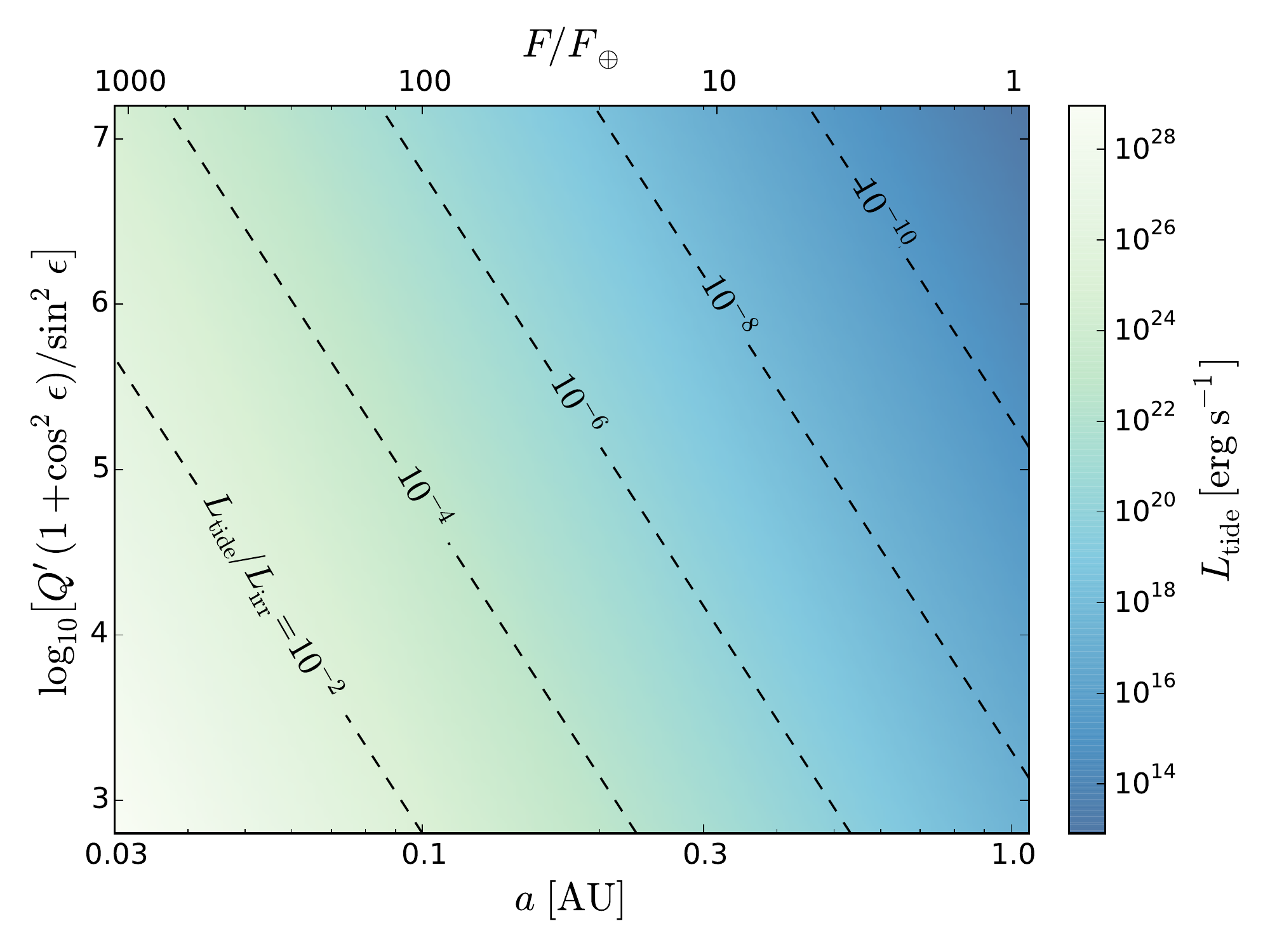}
\caption{Magnitude of obliquity-driven tidal dissipation. Assuming $e=0$ and $\omega = \omega_{\mathrm{eq}}$, we show the dependence of $L_{\mathrm{tide}}$ (equation \ref{e=0 dissipation rate}) as a function of $a$ (bottom x-axis) or $F/F_{\oplus}$ (top x-axis) and $\log_{10}\left[Q^{\prime}(1+\cos^2\epsilon)/\sin^2\epsilon\right]$. The colorbar indicates the magnitude of $L_{\mathrm{tide}}$ in erg s$^{-1}$. The contours show lines of constant $L_{\mathrm{tide}}/L_{\mathrm{irr}}$. We assumed solar values for $R_{\star}$ and $T_{\mathrm{eff}}$, such that $F/F_{\oplus} = (a/{\mathrm{AU}})^{-2}$ and $L_{\mathrm{irr}}/L_{\oplus} = (a/{\mathrm{AU}})^{-2}(R_p/R_{\oplus})^2$. For the planet, we assumed $R_p = 2.5 \ R_{\oplus}$.}
\label{Edot heatmap}
\end{figure}

To get a better handle on the dependence of $L_{\mathrm{tide}}$ on $e$ and $\epsilon$, we plot the normalized tidal dissipation rate, $L_{\mathrm{tide}}/K$, as a heatmap in Figure \ref{Edot_over_K heatmap}. It is clear to see that $L_{\mathrm{tide}}/K$ increases strongly as a function of $e$, and this is especially true for $\epsilon$. Particularly for planets in \textit{Kepler} multiple-transiting systems, which have typical eccentricities of about $e\sim0.05$ \citep{2015ApJ...808..126V, 2019AJ....157..198M}, the dissipation rate at high obliquity will be enhanced by several orders of magnitude compared to the zero obliquity case.

For our implementation of $L_{\mathrm{tide}}$ into the MESA model, we simplify further by taking $e=0$, so as to isolate the effects of obliquity tides.  Equation \ref{dissipation rate} then becomes
\begin{equation}
L_{\mathrm{tide}}(\epsilon) = \frac{9n}{2}\left[\frac{Q'(1+\cos^2\epsilon)}{\sin^2\epsilon}\right]^{-1}\left(\frac{G {M_{\star}}^2}{R_p}\right)\left(\frac{R_p}{a}\right)^6.
\label{e=0 dissipation rate}
\end{equation}
In Figure \ref{Edot heatmap}, we plot the magnitude of $L_{\mathrm{tide}}$ as a function of $a$ (bottom x-axis) or $F/F_{\oplus}$ (top x-axis) and $\log_{10}\left[Q^{\prime}(1+\cos^2\epsilon)/\sin^2\epsilon\right]$. This latter dimensionless quantity summarizes the contribution of the planet parameters (except for $R_p$) to the obliquity-driven tidal heat flux. We also show contours of the ratio of $L_{\mathrm{tide}}$ compared to the incident stellar power, $L_{\mathrm{irr}}$. The ratio $L_{\mathrm{tide}}/L_{\mathrm{irr}}$ approaches a substantial fraction of about $\gtrsim0.01$ for $F/F_{\oplus} \gtrsim 100$. As such, we can anticipate that tides will materially alter planetary structures at least at short orbital periods. 

% May also be useful to compare to the other sources of core luminosity. 
%To understand the degree to which $L_{\mathrm{tide}}$ materially alters the planets' structures, it is instructive to calculate $L_{\mathrm{tide}}$ explicitly and compare it to the magnitude of the incident stellar power, $L_{\mathrm{irr}}$. In terms of the total solar radiation received by Earth, this is given by
%\begin{equation}
%\frac{L_{\mathrm{irr}}}{L_{\oplus}} = \left(\frac{T_{\mathrm{eff}}}{T_{\odot}}\right)^4\left(\frac{a}{\mathrm{AU}}\right)^{-2}\left(\frac{R_p}{R_{\oplus}}\right)^2\left(\frac{R_{\star}}{R_{\odot}}\right)^2.
%\end{equation}

\section{Results}
\label{section 4}

With our thermal evolution model now fully specified, we aim to broadly assess how obliquity tides affect planetary structure over a range of parameter space. Accordingly, we generate $\sim 5000$ models that vary in four principal parameters: the planet mass, $M_p$; the fraction of mass in the H/He envelope, $f_{\mathrm{env}} = M_{\mathrm{env}}/M_p$; the strength of the incident stellar radiation flux with respect to Earth's, $F/F_{\oplus}$; and the strength of the obliquity tides, which is parameterized with $Q'(1+\cos^2\epsilon)/\sin^2\epsilon$ via equation \ref{e=0 dissipation rate}. Each model uses a set of parameters that are uniformly randomly selected within ranges specified in Table \ref{table 2}. We assume solar values for the host star, such that $F/F_{\oplus} = (a/{\mathrm{AU}})^{-2}$. For each set of parameters, we generate two MESA simulations: one that includes tides and one that is tides-free (the latter of which does not depend on the tidal strength parameter). Each simulation is evolved for 10 Gyr.

\begin{table}[h]
\centering
\caption{Parameters and their ranges used for the set of planet models.}
\begin{tabular}{c | c}
Parameter & Range \\
\hline
$M_p/M_{\oplus}$ & (1, 20)    \\
$\log_{10}{f_{\mathrm{env}}}$ & (-2.5, -0.5) \\
%$\log_{10}a$ ($\mathrm{AU}$) & ($\log_{10}0.03$, $\log_{10}0.3$) \\
$\log_{10}F/F_{\oplus}$ & (1, 3) \\
$\log_{10}\left[\frac{Q'(1+\cos^2\epsilon)}{\sin^2\epsilon}\right]$ & (3, 7) \\
%$\log_{10}{Q}$ & (2,5) \\
%$k_2$ & (0.1,0.4) \\
%$\epsilon$ (deg) & (10, 90) \\
\end{tabular}
\label{table 2}
\end{table}

\subsection{Degree of radius inflation}
\label{section 4.1}

% WE NEED TO BETTER CONNECT THE OBS AND THE SIMS. ANSWER THE QUESTION OF WHETHER THE AMOUNT OF RADIUS INFLATION PREDICTED BY THE MODELS IS A CONSISTENT WITH THE DATA. 

\begin{figure}[t!]
\epsscale{1.2}
\plotone{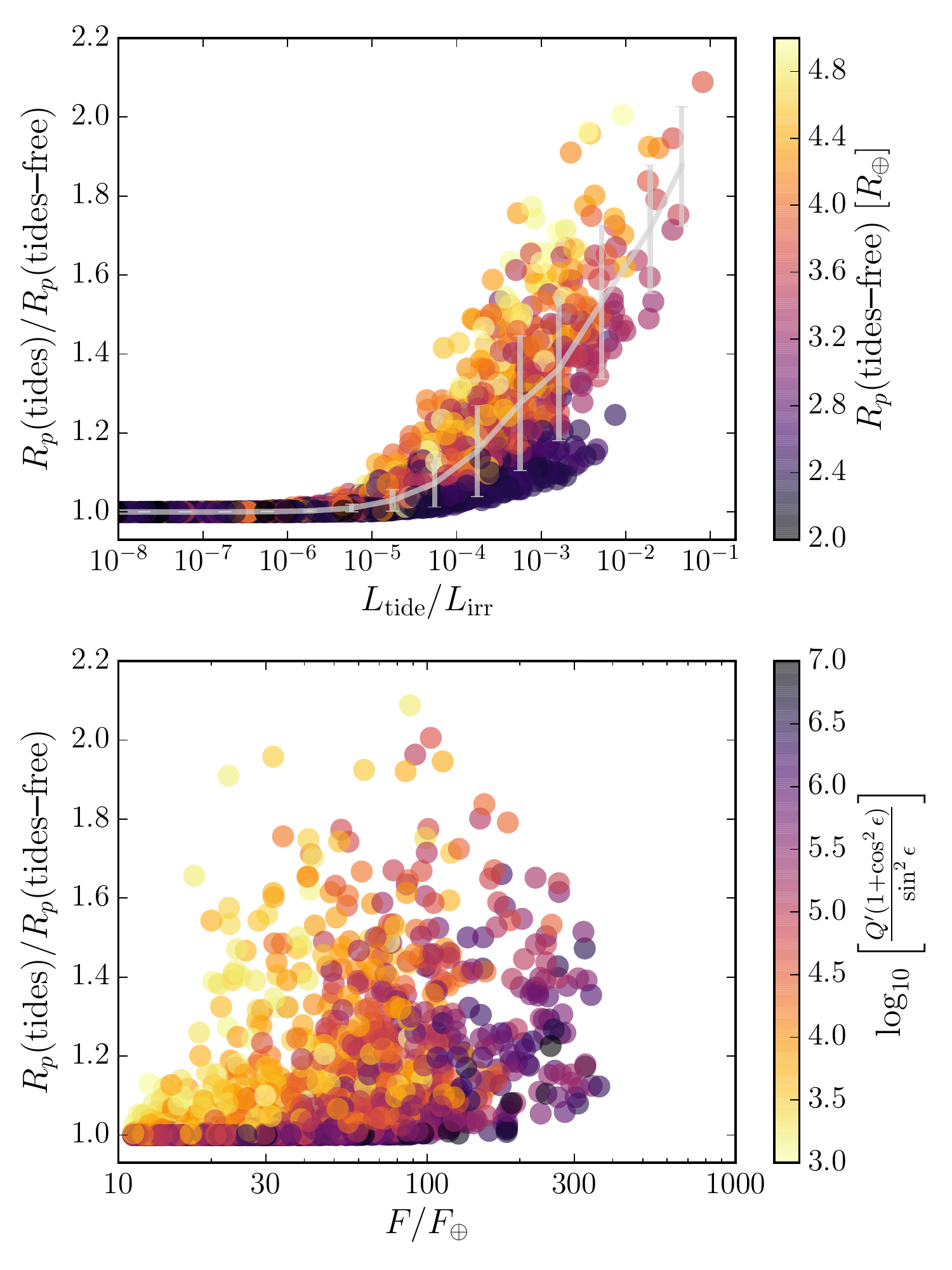}
\caption{Degree of inflation, $R_p$(tides)/$R_p$(tides-free), after 5 Gyr of evolution. The top panel shows this quantity as a function of the fractional tidal heating rate, $L_{\mathrm{tide}}/L_{\mathrm{irr}}$, with coloration corresponding to the radius at 5 Gyr in the tides-free simulation. The gray curve with errorbars shows the mean and standard deviation of the inflation within bins of $L_{\mathrm{tide}}/L_{\mathrm{irr}}$. The bottom panel is a different view using $F/F_{\oplus} = (a/{\mathrm{AU}})^{-2}$ as the x-axis and the tidal efficiency parameter, $\log_{10}\left[Q'(1+\cos^2\epsilon)/\sin^2\epsilon\right]$, as the colorbar. Simulations that went unstable within 10 Gyr have been excluded from these plots.} 
\label{Rp_ratio_vs_Ltide_Lirr}
\end{figure}

Our first goal is to globally quantify the magnitude of tidally-induced radius inflation. To do this, we begin by comparing the evolutionary behavior of planets with and without tides. We extract the planet radii for each simulation at $1, 2, ..., 10$ Gyr and take the ratios of the radii in the tides and tides-free cases, which we shall denote as $R_p$(tides)/$R_p$(tides-free) and define to be the ``degree of inflation''. The top panel of Figure \ref{Rp_ratio_vs_Ltide_Lirr} shows this quantity as a function of the ratio of the tidal luminosity to the incident stellar power, $L_{\mathrm{tide}}/L_{\mathrm{irr}}$, which we also plotted in Figure \ref{Edot heatmap}. The bottom panel decomposes $L_{\mathrm{tide}}/L_{\mathrm{irr}}$ into component parts: $F/F_{\oplus} = (a/{\mathrm{AU}})^{-2}$ (since we're using solar parameters), and $\log_{10}\left[Q'(1+\cos^2\epsilon)/\sin^2\epsilon\right]$.

It is immediately clear from Figure \ref{Rp_ratio_vs_Ltide_Lirr} that tides can quite substantially inflate planetary radii when $L_{\mathrm{tide}}/L_{\mathrm{irr}} \gtrsim 10^{-5}$. In the regime with $L_{\mathrm{tide}}/L_{\mathrm{irr}} \lesssim 10^{-5}$, the extra heat does not perturb the tides-free solution due to the relative dominance of the incident radiation. This is quite similar to the heating threshold for substantial inflation seen in hot Jupiter models \citep[e.g.][]{2007ApJ...661..502B, 2010ApJ...713..751I}. The threshold arises because, in the case of deep heating, the radius deviates from the non-inflated solution when $L_{\mathrm{tide}}\tau_{\mathrm{dep}}/L_{\mathrm{irr}} \gtrsim 10^2$ \citep{2015ApJ...803..111G,2017ApJ...844...94K}, where $\tau_{\mathrm{dep}}$ is the optical depth at which heat is deposited. In our case, the deposition depth is the base of the atmosphere; $\tau_{\mathrm{dep}}$ is dependent on $f_{\mathrm{env}}$ but is usually $\tau_{\mathrm{dep}} \sim 10^7$, which therefore agrees with the threshold occurring at $L_{\mathrm{tide}}/L_{\mathrm{irr}} \gtrsim 10^{-5}$.

\begin{figure*}[t!]
\epsscale{1}
\plotone{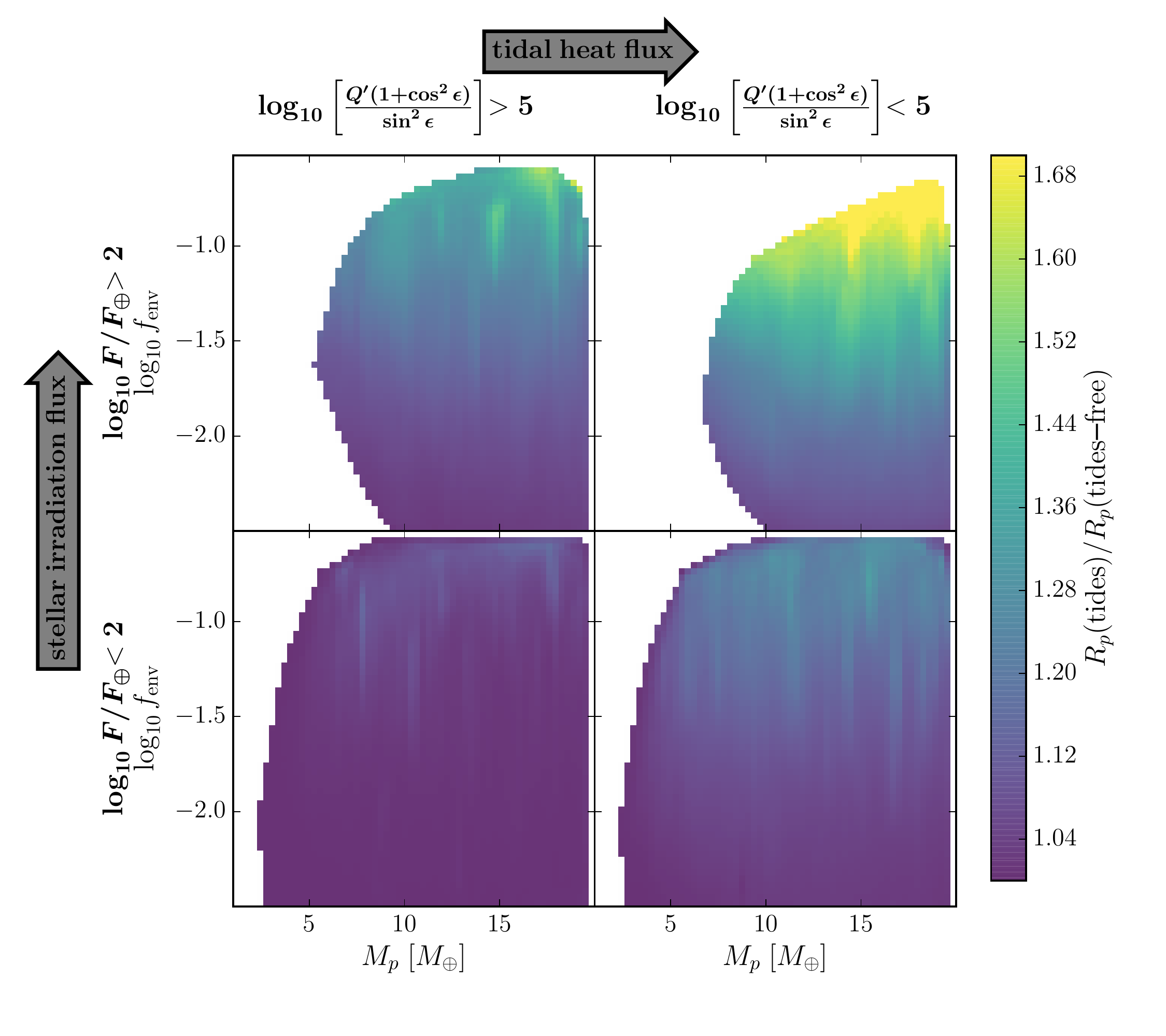}
\caption{Radius inflation behavior throughout parameter space after 5 Gyr of evolution. Each subplot shows the average degree of inflation in the tides and tides-free models as a function of $M_p$ and $\log_{10}f_{\mathrm{env}}$. The radius ratios are calculated from an interpolation to the simulation results and then averaged over ranges in $\log_{10}F/F_{\oplus}$ and
$\log_{10}\left[Q'(1+\cos^2\epsilon)/\sin^2\epsilon\right]$, which are specified in the axis labels. } 
\label{Rp_ratio_grids}
\end{figure*}

For $L_{\mathrm{tide}}/L_{\mathrm{irr}} \gtrsim 10^{-5}$, the inflation degree increases sharply and coherently, and it correlates with the size of the planet (and therefore also $f_{\mathrm{env}}$) in the tides-free case. Typical values are $R_p$(tides)/$R_p$(tides-free)$\sim$1.1--1.5, but planets can inflate up to approximately twice their size in the most extreme cases.

The bottom panel of Figure \ref{Rp_ratio_vs_Ltide_Lirr} demonstrates that substantial radius inflation (up to $\sim100\%$) can occur for any planets with $F/F_{\oplus} \gtrsim 20$ ($a \lesssim 0.22$ AU), as long as the tides are strong enough. The horizontal color gradient and cutoff at $F/F_{\oplus} \approx 400$ indicates that planets with $F/F_{\oplus} \gtrsim 400$ and/or strong tides cannot be stably evolved using these simulations. This is due to rapid atmospheric mass loss, and it will be discussed in greater detail towards the end of this subsection. 

While Figure \ref{Rp_ratio_vs_Ltide_Lirr} provides a sense of the typical radius inflation as a function of the tidal strength, we can gain more insight into this behavior by observing how it varies with planetary properties. To do this, it is helpful to condense and interpolate the simulation results to a high-resolution grid. We define such a grid using the four parameters listed in Table \ref{table 2}. Then, using the $R_p$(tides)/$R_p$(tides-free) ratios at $1, 2, ..., 10$ Gyr, we construct linear barycentric interpolation functions at each evolutionary age and use these functions to calculate the radius ratios at the gridded values.  

Figure \ref{Rp_ratio_grids} summarizes the interpolated radius inflation results over the four parameters at an age of 5 Gyr. Each subplot displays the inflation variation within $M_p$ and $\log_{10}f_{\mathrm{env}}$ space while averaging over ranges (spanning one half of the total ranges) in $\log_{10}F/F_{\oplus}$ and
$\log_{10}\left[Q'(1+\cos^2\epsilon)/\sin^2\epsilon\right]$ space. In this sense, we have divided the results into four regimes and isolated the effects of varying proximity to the star and tidal dissipation independently. 

As expected, inflation is most extreme for close-in planets with strong tides (top right panel of Figure \ref{Rp_ratio_grids}). The degree of inflation can reach up to $R_p$(tides)/$R_p$(tides-free) $=1.7$ in this averaged sense, but, as shown in Figure \ref{Rp_ratio_vs_Ltide_Lirr}, it caps at $\sim$2. The inflation is next largest in the top left panel, the shortest-period planets with slightly lower tidal efficiency, $\log_{10}\left[Q'(1+\cos^2\epsilon)/\sin^2\epsilon\right] > 5$. This is due to the fact that  $L_{\mathrm{tide}} \propto a^{-15/2}$; that is, the tidal dissipation rate is very strongly dependent on the proximity to the star. In all four regimes, the inflation is weakly dependent on $M_p$ but very sensitive to $f_{\mathrm{env}}$. This too makes sense given that sub-Neptune radii in tides-free models are most sensitive to $f_{\mathrm{env}}$ \citep[e.g.][]{2014ApJ...792....1L, 2016ApJ...831..180C}. 

Finally, we note that the white spaces in the subplots of Figure \ref{Rp_ratio_grids} correspond to regimes in which the simulations are unstable. The \cite{2016ApJ...831..180C} MESA model runs into issues at low planet masses, envelope fractions that are either very low or very high, and particularly high levels of irradiation. Low density, highly irradiated planets are more unstable to mass loss driven by photoevaporation, and the timescale for this is too fast for the MESA model to resolve it. Moreover, planets that experience strong tidally-induced radius inflation are more susceptible to mass loss due to their low densities.

However, the near-MMR planets that are relevant to this study are mostly in the $F/F_{\oplus} = 10-100$ range (via inspection of the catalogs from \citealt{2018ApJ...866...99B}). They are therefore not in the regime that is largely affected by the instabilities that occur for $M_p \lesssim 5 \ M_{\oplus}$ in the top panels of Figure \ref{Rp_ratio_grids}. 

% See bottom of Hadden/Lithwick Jupyter notebook to see what to say about the cutoff at 5 Mearth. Also just examine distribution of exoplanets in F/Fearth vs Rp space

Inspection of Figures \ref{Rp_ratio_vs_Ltide_Lirr} and Figure \ref{Rp_ratio_grids} shows that the radius and the degree of inflation are not impacted by all parameters equally and, moreover, that analytic relations may adequately describe the qualitative behavior. Here we use power law models to fit the radii (both with and without tides) and the degree of inflation as a function of $M_p$, $f_{\mathrm{env}}$, $F_p$, and $Q^{\prime}(1+\cos^2\epsilon)/\sin^2\epsilon$. The procedure is similar to that employed by \cite{2014ApJ...792....1L}; \cite{2016ApJ...831..180C} used a quadratic fit instead of a linear one, which resulted in a smaller scatter. Rather than fit the total $R_p$, we fit the envelope contribution to the radius, $R_{\mathrm{env}} = R_p - R_{\mathrm{core}}$. We approximate the core radius as $R_{\mathrm{core}} = 0.97 M_{\mathrm{core}}^{0.28}$ \citep{2016ApJ...831..180C}, which is appropriate for these fits even though the models themselves involved a more complicated relation between $M_{\mathrm{core}}$ and $R_{\mathrm{core}}$. Using least-squares fitting, we derive the following power law relationships:
%No tides
%10^c0 =  1.3140065452948178
%c1 =  -0.2332
%c2 =  0.5627
%c3 =  0.0411
%c4 =  0.0023
%3.4159677469177403
%0.9906869279295613 0.9906739435139873

\begin{equation}
\label{tides-free power law}
\begin{split}
%\frac{R_{\mathrm{env}}\mathrm{(tides\mbox{-}free)}}{R_{\oplus}} &= 
&\frac{R_p\mathrm{(tides\mbox{-}free)} - R_{\mathrm{core}}}{R_{\oplus}} = 1.31\left(\frac{M_p}{10M_{\oplus}}\right)^{-0.23} \\
& \ \  \times\left(\frac{f_{\mathrm{env}}}{0.05}\right)^{0.56}\left(\frac{F_p}{100F_{\oplus}}\right)^{0.041}
%\left(\frac{Q^{\prime}(1+\cos^2\epsilon)/\sin^2\epsilon}{10^5}\right)^{0.0023}
\end{split}
\end{equation}
%Tides
%10^c0 =  1.8965430720679728
%c1 =  -0.2392
%c2 =  0.599
%c3 =  0.2366
%c4 =  -0.041
%11.731420733556625
%0.9729683837880839 0.9729306959299983

\begin{equation}
\label{tides power law}
\begin{split}
%\frac{R_{\mathrm{env}}\mathrm{(tides-free)}}{R_{\oplus}} &= 
&\frac{R_p\mathrm{(tides)} - R_{\mathrm{core}}}{R_{\oplus}} = 1.90\left(\frac{M_p}{10M_{\oplus}}\right)^{-0.24}\left(\frac{f_{\mathrm{env}}}{0.05}\right)^{0.60} \\
&  \ \ \times \left(\frac{F_p}{100F_{\oplus}}\right)^{0.24}\left(\frac{Q^{\prime}(1+\cos^2\epsilon)/\sin^2\epsilon}{10^5}\right)^{-0.041}
\end{split}
\end{equation}

%Radius ratio
%10^c0 =  0.2935570100282113
%c1 =  -0.8956
%c2 =  0.9088
%c3 =  3.1104
%c4 =  -0.8777
%142.46792129982921
%0.9695898779604172 0.9695472569064023
\begin{equation}
\label{deg of inflation power law}
\begin{split}
&\frac{R_p\mathrm{(tides)}}{R_p\mathrm{(tides\mbox{-}free)}} = 1 + 0.29\left(\frac{M_p}{10M_{\oplus}}\right)^{-0.90}\left(\frac{f_{\mathrm{env}}}{0.05}\right)^{0.91} \\
&  \ \ \times \left(\frac{F_p}{100F_{\oplus}}\right)^{3.11}\left(\frac{Q^{\prime}(1+\cos^2\epsilon)/\sin^2\epsilon}{10^5}\right)^{-0.88}
\end{split}
\end{equation}

The $R^2$ values for these fits are 0.99, 0.97, and 0.97. It is also worth noting that our fit for $R_p$(tides-free) is in close agreement with \cite{2014ApJ...792....1L}. Their power law exponents for $M_p$, $f_{\mathrm{env}}$, and $F_p$ were, respectively, -0.21, 0.59, and 0.044. 

These expressions are only an approximation to the simulation results, but they nevertheless illuminate the relative importance of the various parameters. In particular, $F_p$ does not matter much in determining $R_p$(tides-free), but it has a several orders of magnitude stronger influence on $R_p$(tides) and $R_p$(tides)/$R_p$(tides-free). In the latter, $F_p$ is in fact the main contribution, and the rest of the parameters contribute roughly equally with three times smaller power law exponents.

\subsection{Comparison with observations}
\label{section 4.2}
In Section \ref{section 2}, we showed that the mean radius ratios for planets across the 3:2 and 2:1 MMRs are $1.49\pm0.24$ and $2.06\pm0.38$, respectively, or $1.28\pm0.23$ and $1.64\pm0.28$ when restricted to planets with $R_p > 1.8 \ R_{\oplus}$. These latter values are better as comparison metrics, since they involve populations of planets with more significant gaseous components. Broadly speaking, our simulations are consistent with these observational results. Figures \ref{Rp_ratio_vs_Ltide_Lirr} and \ref{Rp_ratio_grids} show that radius inflation in the range $\sim$1.3--2.0 is not only plausible but also common, demonstrating that tidal heating is indeed a possible explanation of the observed radius distribution features. 

In principle, one could attempt a detailed comparison between the observations and models by studying the radii of individual planets wide of the first-order MMRs and comparing them with the set of simulated planets. This is challenging though because each observed planet only has a single radius associated with it rather than a ratio of radii; that is, we have no certain knowledge of the radius in the absence of tides. Moreover, the strong degeneracy between $f_{\mathrm{env}}$ and the tidal efficiency prohibits any strong constraints on either of these parameters. Even though a population-level comparison may not be useful, it is nevertheless interesting to examine at least a few case studies of known exoplanets; we address this in Section \ref{section 5}.

\subsection{Dependence on core composition}
\label{section 4.3}

The simulations presented in the previous sections used a rocky core composition, defined to be 70\% silicates and 30\% Fe. The \cite{2016ApJ...831..180C} model allows for several other core compositions, however, and it is worth examining the sensitivity of the radius inflation with respect to these variations. In this section, we consider another core composition: an ``ice-rock mixture'', which was also explored in \cite{2016ApJ...831..180C} and is defined to be 67\% H$_2$O, 23\% silicates, and 10\% Fe.

\begin{figure}[t!]
\epsscale{1.}
\plotone{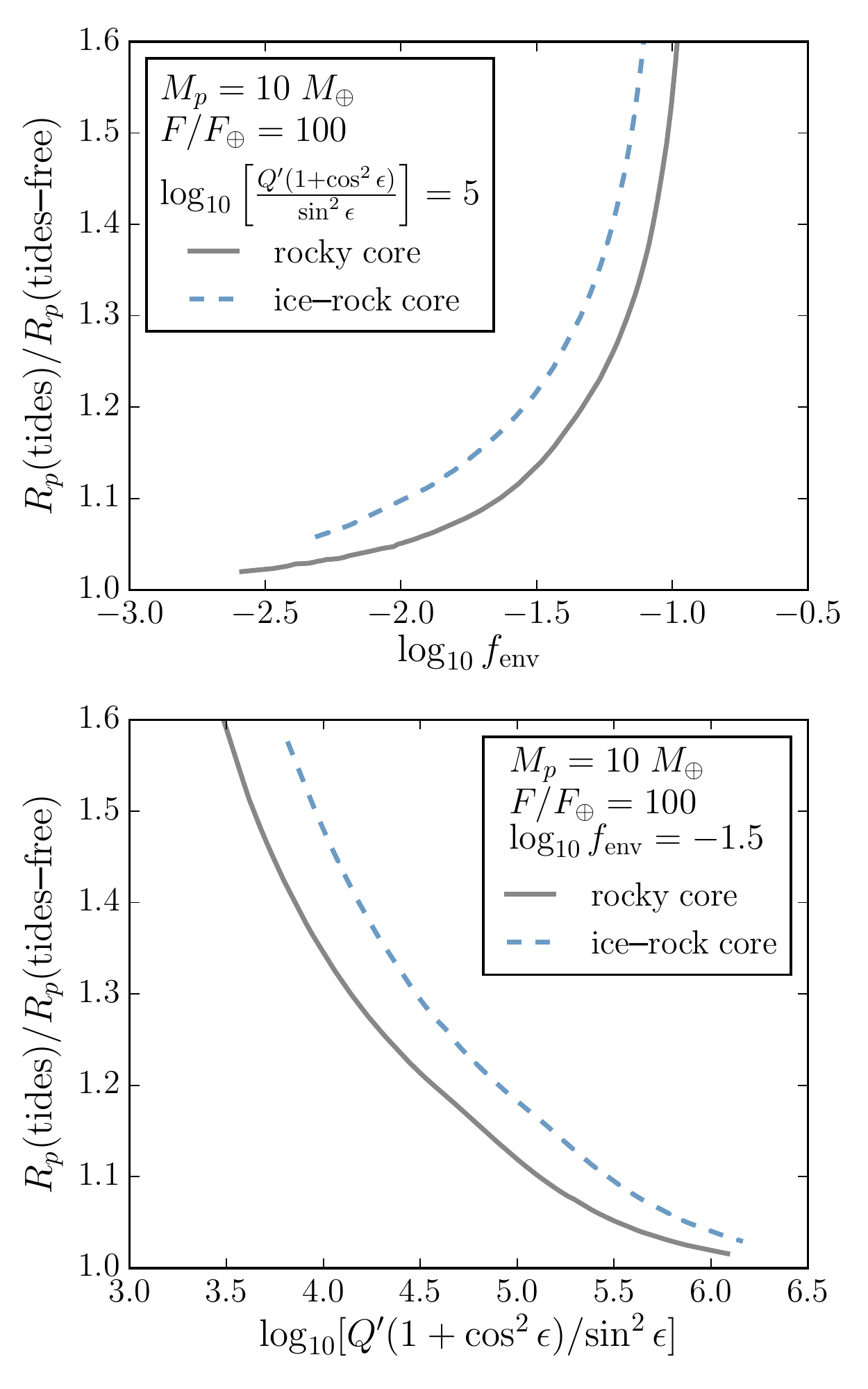}
\caption{Comparisons of the degree of radius inflation for planet models with different core compositions. In both panels, the rocky core model results are plotted in solid gray, and the ice-rock model results are plotted in dashed blue. \textit{Top panel:} Ratio of the radii in the tides and tides-free models as a function of $\log_{10}f_{\mathrm{env}}$ with fixed $M_p$, $F/F_{\oplus}$, and $\log_{10}\left[Q'(1+\cos^2\epsilon)/\sin^2\epsilon\right]$. \textit{Bottom panel:} Radius ratio as a function of $\log_{10}f_{\mathrm{env}}$ with fixed $M_p$, $F/F_{\oplus}$, and $f_{\mathrm{env}}$. } 
\label{Rp_ratio_core_composition}
\end{figure}

We generate a new collection of models by varying either $f_{\mathrm{env}}$ or $Q'(1+\cos^2\epsilon)/\sin^2\epsilon$ and fixing the remaining parameters. Figure \ref{Rp_ratio_core_composition} displays the fixed parameter values and the results of the models after evolution for 2 Gyr. The degree of radius inflation in the ice-rock core models is consistently higher than that in the rocky core models by $\sim 5\%-10\%$. This is an expected result given that ice-rock cores have lower densities and surface gravities than rocky cores of the same mass, so they will be more responsive to tidally-induced heating. However, these lower surface gravities also make planets more susceptible to rapid hydrodynamic mass-loss, which is further exacerbated by the atmospheric inflation. The simulations with strong tides and ice-rock cores tend to undergo instability more frequently.

\subsection{Dependence on deposition depth}
\label{section 4.4}
The tidal heat dissipation has been isolated to the planet's core in all simulations presented thus far. In the MESA models, this is equivalent to depositing it at the base of the H/He envelope. The decision was justified in Section \ref{section 3.1} on the basis of our assessment that the tidal power deposited in the core of a fully-formed sub-Neptune is much more significant than that in the envelope (equation \ref{tidal power ratio}). Even so, it is still important to check how sensitively our results depend on the deposition depth. For instance, the degree of radius inflation in models of hot Jupiters (albeit planets with much deeper atmospheres than sub-Neptunes) is strongly linked to the depth of heating \citep{2015ApJ...803..111G, 2017ApJ...844...94K}. 

We ran an additional set of models with variable $f_{\mathrm{env}}$ but fixed $M_p = 10 \ M_{\oplus}$, $F/F_{\oplus} = 100$, and $\log_{10}\left[Q'(1+\cos^2\epsilon)/\sin^2\epsilon\right] = 5$, as in Section \ref{section 4.3}. For each set of parameters, we ran two simulations: one with the heat deposited in the core as in all of our previous models, and one with the heat deposited uniformly in the convective zone of the atmosphere. Variations of this latter version are often used in hot Jupiter models \citep[e.g.][]{2009ApJ...700.1921I}. 

The radius estimates resulting from these two simulation types are in close agreement overall. The discrepancy is larger for planets with small $f_{\mathrm{env}}$, and it also increases with evolutionary age, but the difference between the radii in the two simulation types is $\lesssim5\%$ and most of the times much smaller than this, $< 0.1\%$. 

We did not run any simulations using even shallower deposition depths, but, given the inflation mechanism discussed in the next section, it is reasonable to expect that very shallow heating at pressures less than $P \lesssim 100$ bars is not capable of inducing substantial radius inflation. The same is true for hot Jupiters \citep{2017ApJ...844...94K}.

We conclude that, although tides are complicated and the heat deposition profile is not known with certainty, our results are not strongly sensitive to these details, as long as the heating occurs at or below the radiative-convective boundary. The radius inflation behavior is therefore robust at least to first order.

\subsection{Mechanism of inflation}

\begin{figure}[t!]
\epsscale{1.2}
\plotone{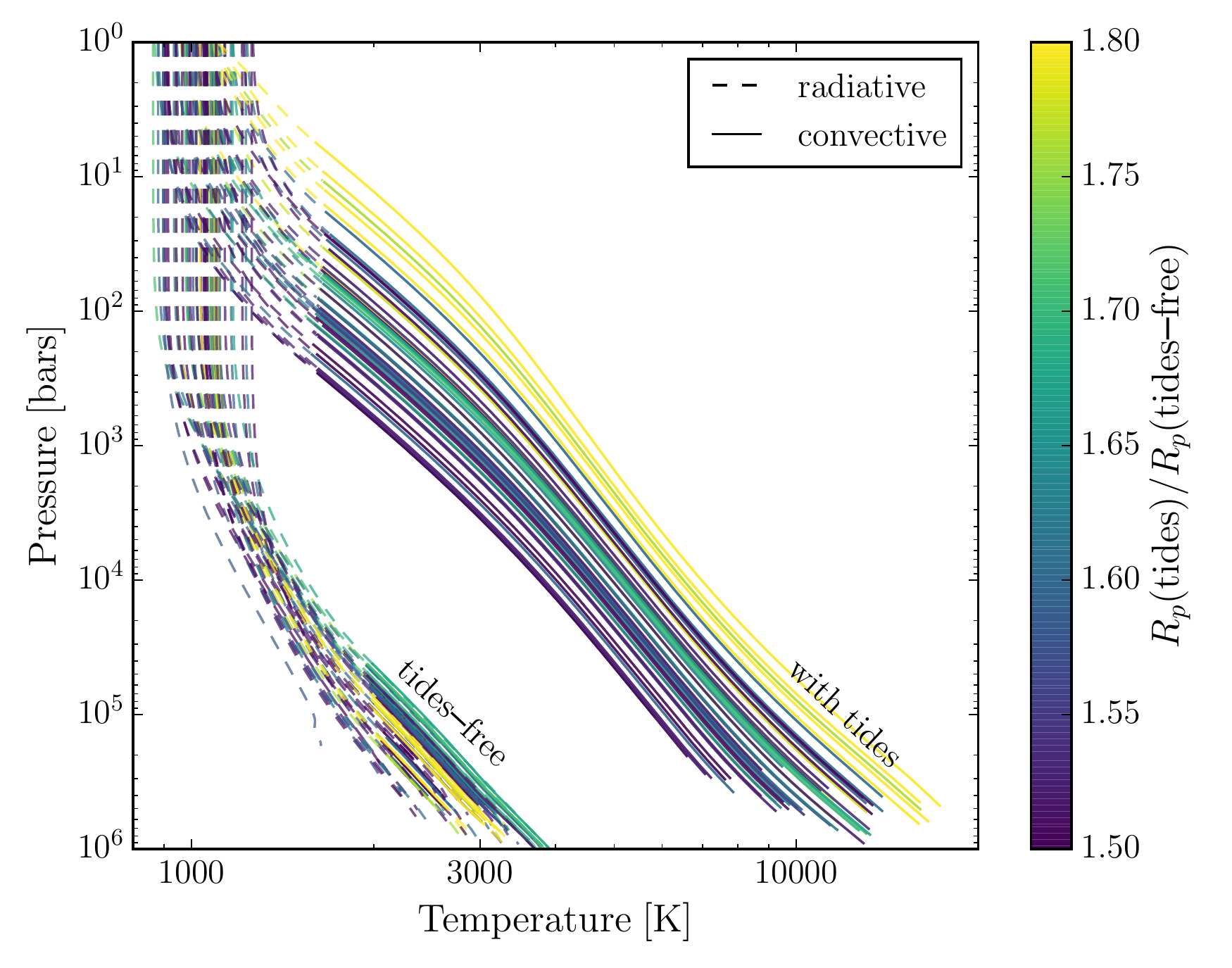}
\caption{Pressure-temperature profiles at 10 Gyr for a set of models with $R_p$(tides)/$R_p$(tides-free)$>1.5$. The left branch of curves are tides-free models, and the right branch are the corresponding models including tides. Within each curve, dashed regions are radiative zones and solid regions are convective. The models with strong tides clearly have steeper profiles for $P \lesssim 100$ bars and much larger convective zones than their tides-free counterparts.} 
\label{P_T_profiles}
\end{figure}

We have shown that tides can induce radius inflation that is $\sim10\%-50\%$ in typical cases and $\sim100\%$ in the most extreme scenarios. However, we have not yet discussed the inflationary mechanisms. It is intuitive that heating at the base of the atmosphere should lead to envelope expansion, but the structural changes that permit this are not immediately clear. To understand this, it is helpful to examine some of the most highly inflated cases. 

Figure \ref{P_T_profiles} shows pressure-temperature profiles at the end state of evolution (10 Gyr) for several models that underwent significant inflation, $R_p$(tides)/$R_p$(tides-free) $>1.5$. The profiles for both the tides and tides-free cases are shown and are colored by the degree of inflation. These profiles show a clear bifurcation. The tides-free model contains a deep outer radiative zone that extends down to $P \gtrsim 10^{5}$ bars, sometimes even lacking a convective region below that. The radiative zone in the tidal model, on the other hand, is quite shallow, extending down to only $P \sim 10-100$ bars. The larger convective zone and steeper adiabatic profile implies that the planet reaches a higher central temperature and entropy, making for an overall larger radius.

\section{Case Studies}
\label{section 5}

% ONE THING TO CONSIDER:  Take the population of ALL planets. Model them with and without tides (I guess that's what we've kind of already done?) and see the radius inflation in a populational sense.

%https://ui.adsabs.harvard.edu/abs/2017MNRAS.466.1868C/abstract
% Kepler-30: https://ui.adsabs.harvard.edu/abs/2018MNRAS.478.2480P/abstract
% Lile & Wang https://iopscience.iop.org/article/10.3847/2041-8213/ab0653/pdf
% Masuda (2014): https://ui.adsabs.harvard.edu/abs/2014ApJ...783...53M/abstract

In the previous sections, we considered a large population of synthetic planets with parameters detailed in Section \ref{section 4} (specifically Table \ref{table 2}), and we studied the behavior of their tidally-driven radius inflation. This has been instructive, but it is worthwhile to extend our examination to known exoplanets. In this section, we identify and characterize several planets that are intriguing case studies for potentially experiencing radius inflation.

One subset of known sub-Neptunes immediately comes to mind. This is the population of ``super-puffs'', or planets with anomalously low bulk densities, $\rho \lesssim 0.1 \ \mathrm{g \ cm^{-3}}$. Examples include planets in the Kepler-51 \citep{2014ApJ...783...53M}, Kepler-79 \citep{2014ApJ...785...15J}, and  Kepler-87 \citep{2014A&A...561A.103O} systems. In order to explain their densities, super-puffs are typically inferred to contain large gas fractions \citep[e.g.][]{2016ApJ...831..180C, 2017AJ....154....5H}, but it has been a challenge to explain how they could accrete and retain these envelopes and also avoid runaway gas accretion that would otherwise turn them into gas giants \citep[e.g.][]{2016ApJ...817...90L, 2017MNRAS.464.3937G, 2019ApJ...873L...1W}. It is possible that some or all of these super-puffs do not actually have such high envelope fractions; rather, their low densities could be generated by strong tidally-induced radius inflation. The population of super-puffs is therefore a good place to start for the basis of our case studies. We also, however, need to find which of these planets have orbital period ratios just wide of the first-order MMRs.

We begin our search for case studies using a sample of 145 planets in 55 systems with TTV-measured masses from \cite{2017AJ....154....5H}. We calculate the period ratios of all pairs of planets within these systems and extract those with period ratios in the ranges $(1.5, 1.56)$ and $(2.0, 2.07)$, i.e. just wide of the 3:2 and 2:1 MMRs. Next, we limit the sample to planets with $2 \ M_{\oplus} < M_p < 20 \ M_{\oplus}$ and $10 < F/F_{\oplus} < 1000$, such that they fall within the parameter domain of our models from Section \ref{section 4}. Finally, we limit the search to planets with densities, $\rho \leq 0.1 \ \mathrm{g \ cm^{-3}}$. Planets with densities greater than this threshold could certainly also be tidally inflated. However, for our purposes here, it is useful to examine the most extreme cases.

%Finally, we search for planets that might show evidence of radius inflation. This is largely ill-defined prior to any modeling efforts, but radius inflation might be suggested by any planets that substantially deviate from typical $\lesssim 4 \ R_{\oplus}$ sub-Neptune radii or break the observed intra-system size uniformity that compact multi-planet systems exhibit  \citep{2017ApJ...849L..33M, 2018AJ....155...48W}. As a first pass, we impose a very simple criterion that the planet's radius is larger than the minimum radius in the system. 

Using this procedure, we identify 
%39 planets in 15 systems as potential case studies. 
5 planets in 3 systems as potential case studies: Kepler-27 b and c, Kepler-31 b and c, and Kepler-79 d. Kepler-51 c and d would have also been included, but their irradiation fluxes are too small for the parameter domain of our models. We profile these case studies, as well as some of the others in their systems, in the subsequent sections. We begin with the Kepler-79 system because we think it is the most interesting, and we then proceed to the Kepler-31 and Kepler-27 systems. 
%A comprehensive analysis of all candidates would be worthwhile but is beyond the scope of this work. 

\subsection{The Kepler-79 planets}

\begin{figure}[t!]
\epsscale{1.2}
\plotone{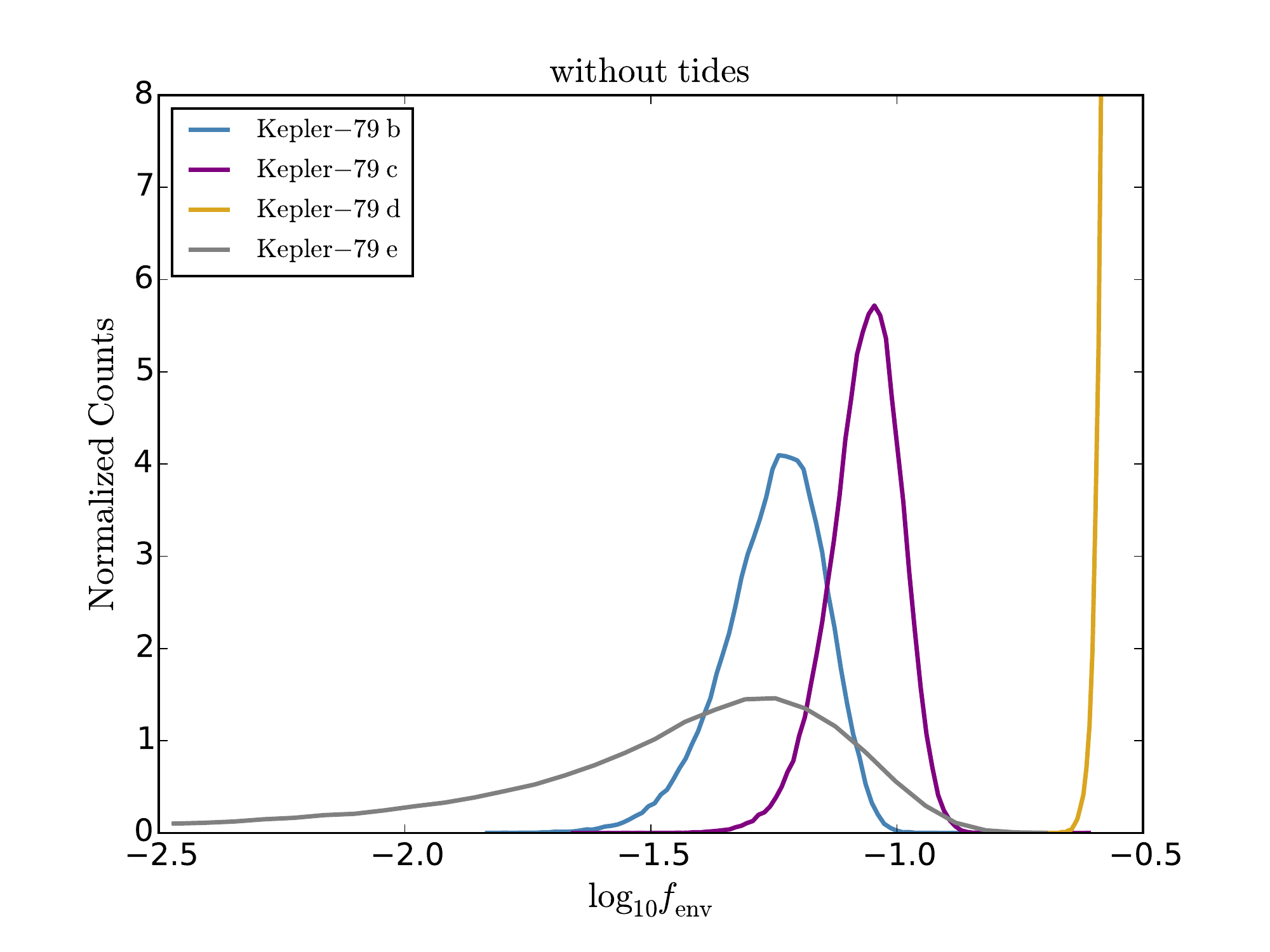}
\caption{Estimates of the envelope fractions, $f_{\mathrm{env}}$, of the Kepler-79 planets in a tides-free model. This plot shows marginalized posterior distributions of the tides-free fit described in \ref{section 5}. (Note: The y-axis is truncated for visibility purposes, but planet d's curve keeps rising towards the right.) The distribution means and uncertainties (calculated using the 16th and 84th percentiles) are as follows. Planet b: $f_{\mathrm{env}} = 5.6^{+1.3}_{-1.3}\%$; planet c: $f_{\mathrm{env}} = 8.6^{+1.4}_{-1.4}\%$; planet d: $f_{\mathrm{env}}\gtrsim 30\%$, planet e: $f_{\mathrm{env}} = 4.5^{+2.8}_{-2.9}\%$. }
\label{Kepler-79 without tides}
\end{figure}

\begin{figure*}[t!]
\epsscale{1}
\plotone{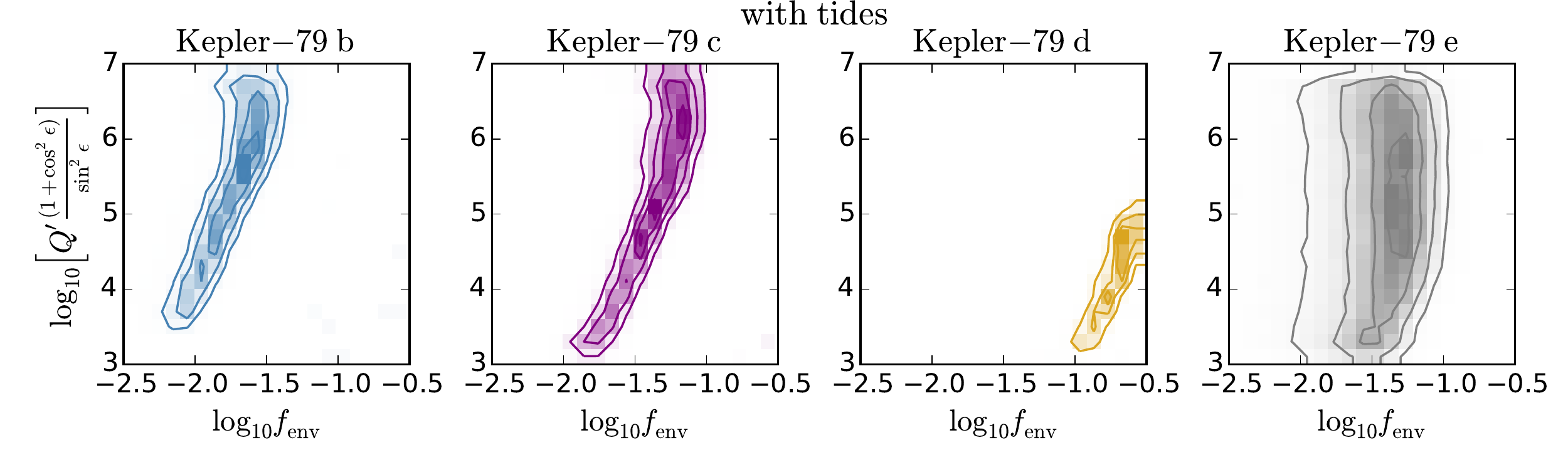}
\caption{Estimates of $\log_{10}[Q'(1+\cos^2\epsilon)/\sin^2\epsilon]$ and $\log_{10}f_{\mathrm{env}}$ of the Kepler-79 planets in a model that includes tides. Each plot shows the density and contours of the 2D posterior distributions.} 
\label{Kepler-79 with tides}
\end{figure*}

% Jontof-Hutter 2014 https://iopscience.iop.org/article/10.1088/0004-637X/785/1/15/pdf

Kepler-79 is an anomalous system with four, very low density planets \citep{2014ApJ...785...15J}. The host star -- which has properties, $M_{\star} = 1.25^{+0.04}_{-0.04} \ M_{\odot}$, $R_{\star} = 1.30^{+0.15}_{-0.10} \ R_{\odot}$, $T_{\mathrm{eff}} = 6388^{+63}_{-66}$ K \citep{2017AJ....154..107P, 2017AJ....154..108J} -- is orbited by a compact configuration of planets with period ratios near a 1:2:4:6 chain of commensurability. The planetary properties are listed in Table \ref{Kepler-79 parameters}, and the adjacent pair period ratios are $P_c/P_b = 2.032$, $P_d/P_c = 1.901$, $P_e/P_d = 1.556$. The CKS team derived the system age to be $1.91^{+0.81}_{-1.05}$ Gyr. This is slightly lower but consistent with the findings of \cite{2014ApJ...785...15J}, who found the age to be $3.44^{+0.60}_{-0.91}$ Gyr. 

\begin{table}[h]
\centering
\caption{Parameters of the Kepler-79 system.\footnote{Orbital periods are obtained from the \textit{Kepler} DR 25 KOI Catalog \citep{2018ApJS..235...38T}; $F/F_{\oplus}$ values and radii are from a \textit{Gaia} DR2 analysis \citep{2018ApJ...866...99B}; masses are TTV-derived \citep{2017AJ....154....5H}.}}
\begin{tabular}{c | c c c c }
Planet & $P$ [days] & $F/F_{\oplus}$ & $M_p \ [M_{\oplus}]$ & $R_p \ [R_{\oplus}]$ \\
\hline
Kepler-79 b & 13.485 & 205.4$^{+9.4}_{-9.0}$ & 8.3$^{+7.6}_{-3.7}$ & 3.44$^{+0.17}_{-0.16}$ \\
Kepler-79 c & 27.402 & 79.8$^{+3.6}_{-3.5}$ & 6.6$^{+3.4}_{-2.0}$ &  3.68$^{+0.18}_{-0.17}$ \\
Kepler-79 d & 52.091 & 33.9$^{+1.5}_{-1.5}$ & 6.6$^{+1.9}_{-2.0}$ &   7.24$^{+0.34}_{-0.32}$ \\
Kepler-79 e & 81.064 & 18.8$^{+0.9}_{-0.8}$ & 3.9$^{+1.1}_{-0.9}$ &   3.05$^{+0.76}_{-0.19}$ \\  
\end{tabular}
\label{Kepler-79 parameters}
\end{table}

Planet d has the most anomalous radius for its mass, but the others also warrant exploration. In order to examine whether planet d or any of the others plausibly underwent radius inflation, we fit all planets using both tidal and tides-free models. Our procedure is as follows. Similar to Section \ref{section 4}, we employ the simulation set to construct linear barycentric interpolation functions for $R_p$(tides) and $R_p$(tides-free) as a function of $M_p$, $\log_{10}f_{\mathrm{env}}$, $\log_{10}(F/F_{\oplus})$, and $\log_{10}\left[Q'(1+\cos^2\epsilon)/\sin^2\epsilon\right]$. These interpolation functions are specified at an evolutionary age of 2 Gyr so as to maintain consistency with system's estimated age. We also tested the methods using other ages and did not find strong sensitivity with respect to this. 

Using these interpolation functions as radius models, we employ a Markov Chain Monte Carlo (MCMC) method to estimate the parameters consistent with the observed radii. We use the affine invariant ensemble sampler \texttt{emcee} \citep{2010CAMCS...5...65G, 2013PASP..125..306F} using 200 walkers. For both models (with and without tides), we fix $F/F_{\oplus}$ to the observed values for each planet and let $M_p$ float within uncertainties. As such, the only important free parameter in the tides-free model is $\log_{10}f_{\mathrm{env}}$. The obliquity tide strength $\log_{10}\left[Q'(1+\cos^2\epsilon)/\sin^2\epsilon\right]$ is an additional free parameter in the tidal model. Using uniform priors and a Gaussian likelihood function, we collect 5000 MCMC samples, discard the first 1000 as burn-in, and use the remaining posterior samples for parameter estimation.

Figure \ref{Kepler-79 without tides} shows the posterior distributions of $\log_{10}f_{\mathrm{env}}$ according to the tides-free model. The $f_{\mathrm{env}}$ estimates for planets b, c, d, and e are, respectively, $5.6^{+1.3}_{-1.3}\%$, $8.6^{+1.4}_{-1.4}\%$, $\gtrsim 30\%$, and $4.5^{+2.8}_{-2.9}\%$. Even with updated observed radius measurements, our estimates are generally consistent with \cite{2014ApJ...792....1L}, who found $6.6^{+0.7}_{-1.0}\%$, $8.9^{+0.7}_{-0.9}\%$, and $36.7^{+3.6}_{-3.4}\%$, $8.0^{+1.1}_{-1.1}\%$ for planets b, c, d, and e. It is clear to see that planet d is an outlier among the others in the system, which is unusual given typical intra-system uniformity of the \textit{Kepler} multis \citep{2017ApJ...849L..33M, 2018AJ....155...48W}. As we now show, planet d's $f_{\mathrm{env}}$ might not actually be that large, but rather just appear to be because it has been tidally inflated.

Figure \ref{Kepler-79 with tides} shows the 2D posterior distributions resulting from the tidal model fit. We clearly see the expected degeneracy between the tidal parameters and $f_{\mathrm{env}}$; the degeneracy weakens with increasing orbital period (since tides also weaken with increasing period). The case for tidal inflation is most compelling for planet d, which would require tidal parameters $\log_{10}\left[Q'(1+\cos^2\epsilon)/\sin^2\epsilon\right] \sim 3-4$ to reach $f_{\mathrm{env}} \sim 10\%$, a value that is much closer to the other planets in the system and the broader population of sub-Neptunes \citep[e.g.][]{2014ApJ...792....1L, 2016ApJ...831..180C}. Figure \ref{Kepler-79 with tides} shows that planets b, c, and e could also be affected by tides, perhaps with smaller obliquities or different tidal parameters.

\subsection{Kepler-31 c}

The Kepler-31 system -- whose parameters are shown in Table \ref{Kepler-31 parameters} -- is intriguing because the radii of the three confirmed planets decrease with increasing orbital period, which is opposite the typical size-ordering trend \citep{2018AJ....155...48W, 2018MNRAS.473..784K}. The three outer planets orbit their $M_{\star} = 1.12^{+0.08}_{-0.06} \ M_{\odot}$ \citep{2017AJ....154..107P, 2017AJ....154..108J} host star in a near-4:2:1 resonant chain, with period ratios $P_c/P_b = 2.044$ and $P_d/P_c = 2.056$. The system is estimated to be 4.7$^{+0.9}_{-0.8}$ Gyr old \citep{2017AJ....154..107P, 2017AJ....154..108J}. 

\begin{table}[h]
\centering
\caption{Parameters of the Kepler-31 system. (See Table \ref{Kepler-79 parameters} footnote.)}
\begin{tabular}{c | c c c c }
Planet & $P$ [days] & $F/F_{\oplus}$ & $M_p \ [M_{\oplus}]$ & $R_p \ [R_{\oplus}]$ \\
\hline
KOI-935.04 & 9.617 & 256.5$^{+24.2}_{-21.8}$ & -- & 1.95$^{+0.14}_{-0.13}$ \\
Kepler-31 b & 20.860 & 91.5$^{+8.6}_{-7.8}$ & 0.7$^{+2.4}_{-0.6}$ & 5.51$^{+0.35}_{-0.32}$ \\
Kepler-31 c & 42.634 & 35.3$^{+3.3}_{-3.0}$ & 2.2$^{+4.6}_{-1.2}$ & 5.23$^{+0.33}_{-0.31}$ \\ 
Kepler-31 d & 87.647 & 13.5$^{+1.3}_{-1.1}$ & 2.8$^{+4.2}_{-1.4}$ & 4.01$^{+0.27}_{-0.24}$ \\ 
\end{tabular}
\label{Kepler-31 parameters}
\end{table}

\begin{figure}[t!]
\epsscale{1}
\plotone{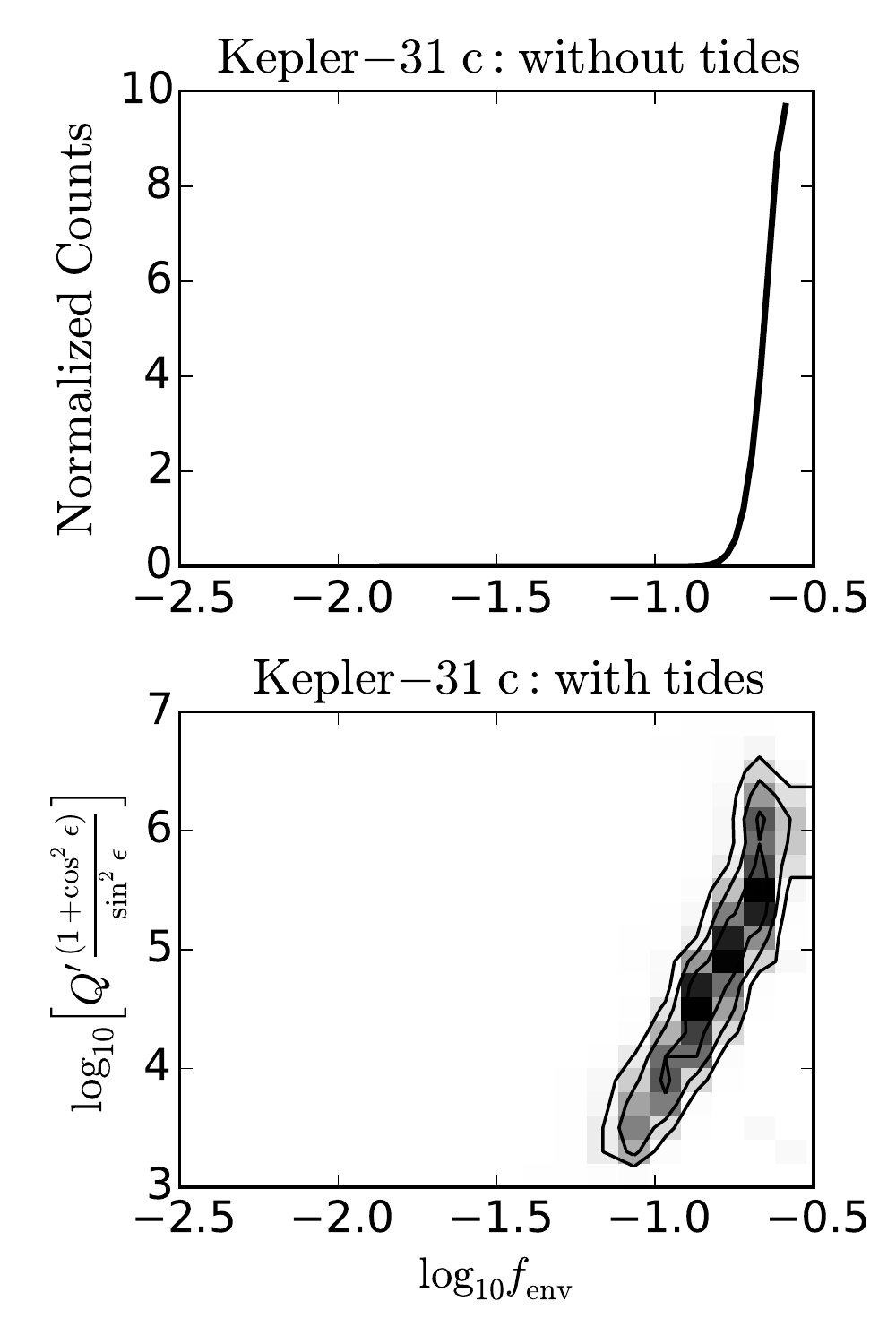}
\caption{Posterior distributions of the parameters of Kepler-31 c according to the two models with and without tides. The top panel shows the distribution of $\log_{10}f_{\mathrm{env}}$ in the tides-free model. The envelope fraction estimate is $f_{\mathrm{env}}\gtrsim 30\%$. The bottom panel shows the density and contours of posterior distributions of $\log_{10}[Q'(1+\cos^2\epsilon)/\sin^2\epsilon]$ and $\log_{10}f_{\mathrm{env}}$ in the tidal model.}
\label{Kepler-31}
\end{figure}

We apply similar techniques as we did for Kepler-79 to investigate the possibility of radius inflation in this system. Planet b is a strong candidate based on its observed parameters, but its mass makes it too small to study using our simulation set. Planet c is the next best candidate given its radius. In Figure \ref{Kepler-31}, we display the posterior distributions resulting from the two analyses with and without tides. According to the tides-free model, the envelope fraction is $f_{\mathrm{env}}\gtrsim 30\%$. If there is tidal inflation, however, the envelope fraction could be much more typical, as low as $\sim 5\%$.

\newpage
\subsection{Kepler-27 b}
Kepler-27 is similarly interesting to Kepler-31. The $4.7^{+5.1}_{-3.3}$ Gyr old system consists of four planets (two confirmed, two candidate) orbiting a $0.89^{+0.03}_{-0.03} \ M_{\odot}$ star \citep{2017AJ....154..107P, 2017AJ....154..108J}. The parameters of the inner three planets are shown in Table \ref{Kepler-27 parameters}. Planets b and c are wide of the 2:1 MMR, with a period ratio equal to $P_c/P_b = 2.043$. Both planets appear to be strong candidates for inflation given their large radii/small densities. 

\begin{table}[h]
\centering
\caption{Parameters of the inner three planets in the Kepler-27 system. (See Table \ref{Kepler-79 parameters} footnote.)}
\begin{tabular}{c | c c c c }
Planet & $P$ [days] & $F/F_{\oplus}$ & $M_p \ [M_{\oplus}]$ & $R_p \ [R_{\oplus}]$ \\
\hline
KOI-841.03 & 6.546 & 89.3$^{+9.1}_{-8.1}$ & -- & 2.04$^{+0.50}_{-0.18}$ \\
Kepler-27 b & 15.335 & 28.7$^{+2.9}_{-2.6}$ & 3.6$^{+1.7}_{-1.6}$ & 4.65$^{+0.31}_{-0.52}$ \\
Kepler-27 c & 31.331 & 11.1$^{+1.1}_{-1.0}$ & 4.3$^{+ 2.0}_{-1.6}$ & 6.06$^{+0.40}_{-0.40}$ \\
\end{tabular}
\label{Kepler-27 parameters}
\end{table}

\begin{figure}[t!]
\epsscale{1}
\plotone{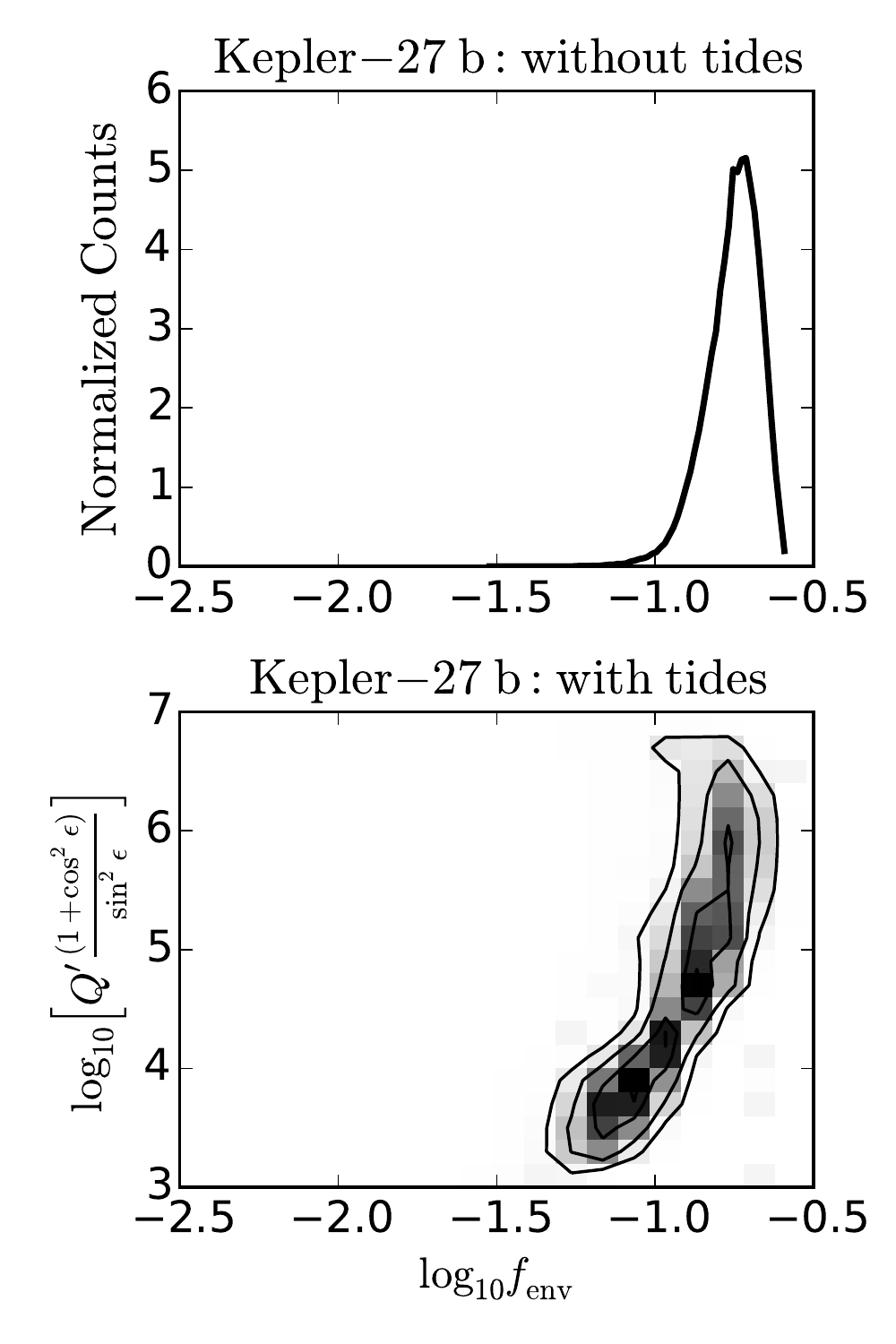}
\caption{Same as Figure \ref{Kepler-31}, but for Kepler-27 b. According to the tides-free model, the envelope fraction estimate is $f_{\mathrm{env}} = 17.7^{+3.4}_{-3.3}\%$.}
\label{Kepler-27}
\end{figure}

We apply similar techniques as we did for the Kepler-79 and Kepler-31 systems. Planet c's stellar irradiation flux is too close to the lower bound of our simulation set's range for satisfactory modeling. Accordingly, we only present results for planet b. In Figure \ref{Kepler-27}, we display the posterior distributions resulting from the two analyses with and without tides. According to the tides-free model, the envelope fraction is $f_{\mathrm{env}} = 17.9^{+3.3}_{-3.4}\%$. If there is tidal inflation, however, the envelope fraction could be as low as $\sim 3\%-5\%$.

\section{Discussion}
\label{section 6}

\subsection{Discrepancy between the TTV and RV populations}

In Section \ref{section 2} (and preliminarily shown in \citealt{2019NatAs...3..424M}), we found that planets in pairs wide of first-order MMRs have average radii that are statistically significantly larger than planets at other nearby period ratios. These radius distribution features have important implications for the observed discrepancy between the populations of planets with TTV-measured masses and RV-measured masses. A number of authors have noted that TTV planets appear to have systematically smaller densities than RV planets \citep{2014ApJ...783L...6W, 2014ApJ...785...15J, 2016MNRAS.457.4384S, 2016ApJ...825...19W, 2017ApJ...839L...8M}. This is seen as a $\sim 2\sigma$ offset between the two populations in the mass-radius diagram \citep{2016ApJ...825...19W}. 

Observational and sensitivity biases are paramount to a full understanding this discrepancy \citep{2016MNRAS.457.4384S, 2017ApJ...839L...8M}. Specifically, the sensitivity of the TTV and RV methods depends upon the planetary physical and orbital properties; TTV sensitivity increases with increasing planet radius and orbital period, while RV sensitivity decreases with increasing orbital period. For a given mass, RV planets will be on systematically smaller orbits (and thereby also consist of systematically denser planets), and for a given radius, TTVs will be capable of detecting planets with systematically lower masses \citep{2016MNRAS.457.4384S, 2017ApJ...839L...8M}. 

Before now, it was unclear whether these biases were sufficient to fully explain the density discrepancy between the TTV and RV populations. However, in this work, we have shown that there are indeed astrophysical (i.e. intrinsic) differences in planets just wide of MMR; they have larger radii on average. Near-MMR proximity is required for most TTV measurements because these orbital configurations yield the strongest TTV signals \citep[e.g.][]{2017AJ....154....5H}. As a result, TTV measurements are biased to near-MMR planets, which have intrinsically larger radii and smaller densities. Both bias and astrophysical differences must explain the TTV/RV population discrepancy; further study is required to deduce the relative importance of these effects.   

%https://ui.adsabs.harvard.edu/abs/2016MNRAS.457.4384S/abstract
%https://iopscience.iop.org/article/10.3847/2041-8213/aa67eb/pdf
%Steffen: Rather, these differences largely stem from the fact that the sensitivity of the two methods depends upon the physical and orbital properties of the planets themselves. This ‘sensitivity bias’ means that the two methods are better suited to probing different regions on the mass–radius plane.... Comparing equations (14) and (3) we see that a large planet radius or a longer orbital period can enable measurements of planets with smaller masses. The different dependencies on orbital period imply that, for a given mass, RV planets will have systematically shorter orbits.... The difference in SNR (i.e. the sensitivity bias) for the two methods is partly responsible for the different densities of planets measured by them – rather than incorrect or biased mass measurements themselves.... We see that for a given planet size, TTVs are able to detect planets with lower masses. This fact is responsible, at least partially, for the observation that planets characterized by TTVs have had systematically lower density than those measured by RVs.

\subsection{Future modeling considerations}
Simplifying assumptions must be made when studying a system as inherently complex as a planet's interior structure and atmosphere, and our study has made several such simplifications. For instance, our model consisted of a spherically symmetric 1D atmosphere, despite the fact that tidal distortion is by definition a 3D phenomenon. We assumed a H/He-dominated atmosphere and used default atmospheric metallicities and opacities from \cite{2016ApJ...831..180C}. We did not consider variations in stellar spectral type. We also did not consider any variations with respect to the time at which the tides become active. These simplifications are appropriate for our attempt at first-order understanding, but in future work, it would be useful to explore more diverse conditions.

The atmospheric metallicity is a particularly interesting parameter for future exploration. High metallicities have been inferred for some Neptune and sub-Neptune-mass planets \citep[e.g.][]{2017AJ....153...86M}. The increased opacity at high metallicity will result in the radiative-convective boundary moving to lower pressures, thereby widening the convective zone. This may increase the efficiency of the tidally-induced radius inflation mechanism, such that the degree of radius inflation is larger for the same amount of heating.

Another interesting direction for future work is the exploration of diverse interior and atmosphere compositions. The densities of many sub-Neptunes cannot uniquely distinguish them between ``gas dwarfs'' and ``water worlds'' \citep[e.g.][]{2008ApJ...673.1160A}. Although most auxiliary signatures in the planet population point towards the gas dwarf interpretation, it would be interesting to investigate whether the observed radius enhancement features can rule out the steam atmospheres of water worlds, if the features are indeed a result of tidal heating.

\section{Summary}
\label{section 7}
This work was motivated by an intriguing trend: Planets with period ratios wide of first-order MMRs have enhanced radii compared to planets at other period ratios. More specifically, planets just wide of the 2:1 and 3:2 orbital period ratios are, respectively, $2.06\pm0.38$ and $1.49\pm0.24$ times larger than planets just inside these resonances. 

Previous work suggested that these wide-of-MMR planets should have preferentially non-zero obliquities (axial tilts) in order to fully explain the pile-up wide of MMR in the first place. In these high obliquity states, the tidal dissipation that the planets experience is much stronger, inevitably resulting in more heat energy being deposited in the planets. In this paper, we asked the following question: Is the heat from these ``obliquity tides'' strong enough to inflate the planets' radii, providing a consistent explanation for the observed radius trends? Even if this is not the case, we were also motivated by a broader question: How does the structure of a sub-Neptune planet respond to strong tides driven by excited eccentricities or obliquities?  

To address these questions, we adapted a robust thermal evolution model developed for sub-Neptunes by \cite{2016ApJ...831..180C} using the Modules for Experiments in Stellar Astrophysics (MESA) toolkit. We added an extra source of interior heat presumed to arise from obliquity tides. We deposited this heat in the core, based on an argument that dissipation in the core should dominate over dissipation in the envelope, but our results vary little in any case where the heat is deposited at or below the radiative-convective boundary. 

Planets experience significant structural changes when the ratio of the tidal luminosity to the incident stellar power, $L_{\mathrm{tide}}/L_{\mathrm{irr}} \gtrsim 10^{-5}$. This tidal power is common for planets on short orbits, $a \lesssim 0.3$ AU, with reasonable tidal quality factors and obliquities, $\log_{10}\left[Q^{\prime}(1+\cos^2\epsilon)/\sin^2\epsilon\right] \sim 3-7$. In the $L_{\mathrm{tide}}/L_{\mathrm{irr}} \gtrsim 10^{-5}$ regime, the degree of radius inflation is $R_p$(tides)/$R_p$(tides-free)$\sim 1.1-1.5$ in typical cases, but planets can expand by over two times their size in extreme cases. The model radii and degree of radius inflation are well-described by power laws of planetary parameters (equations \ref{tides-free power law}--\ref{deg of inflation power law}). Overall, the results are robust with respect to changes in the core composition and the heat deposition depth. The radius inflation is accompanied by a large expansion in the size of the convection zone.

In the final section, we examined three systems containing ``super-puff'' planets with very low densities: the Kepler-79 planets (particularly planet d), Kepler-31 c, and Kepler-27 b. These planets are strong candidates for tidally-induced radius inflation. In the absence of tides, all planets would require envelope mass fractions $f_{\mathrm{env}}\sim 30\%$, which is significantly greater than the typical envelope fractions for sub-Neptunes ($\sim1\%-10\%$). Accounting for tides, the extra heating can generate a large part of these planets' distended radii, thereby obviating the necessity for large envelopes and rather only requiring $f_{\mathrm{env}} \sim 3\%-10\%$. 

This work to identify and characterize potential signatures of tidal inflation in the sub-Neptune population is still early, and it is expected to improve with further observations and modeling efforts. NASA's Transiting Exoplanet Survey Satellite (TESS) Mission \citep{2015JATIS...1a4003R} will augment the statistics of this population by allowing radii and masses for additional planets in near-resonant pairs to be measured. In turn, thermal evolution models that include tidal heating will allow us to glean deeper insights into the structures and compositions of these fascinating worlds.

%**Inhibition of contraction**

\section{Acknowledgements}
I am grateful to Greg Laughlin for inspiring conversations, as well as Howard Chen, Chris Spalding and Dong Lai for useful comments. Thank you also to Howard Chen \& Leslie Rogers for their publicly available MESA model, as well as Thaddeus Komacek for his assistance in updating this model to the latest MESA version. Finally, we thank the referee, Eric Lopez, for insightful comments. This material is based upon work supported by the NSF Graduate Research Fellowship Program under Grant  DGE-1122492. This research has made use of the NASA Exoplanet Archive, which is operated by the California Institute of Technology, under contract with the National Aeronautics and Space Administration under the Exoplanet Exploration Program.

\bibliographystyle{aasjournal}
\bibliography{main}

\end{document}